\documentclass[usenatbib]{mnras}
\usepackage{epsfig}
\usepackage{natbib}
\usepackage{latexsym}
\usepackage{amsmath}
\usepackage{setspace}

\newcommand{\degrees}{$^\circ$}
\newcommand{\etal}{et~al.} 
\newcommand{\ionhy}{H{\sc ii} }

\newcommand{\kms}{$\mbox{km~s}^{-1}$}
\newcommand{\msol}{\mbox{M\hbox{$_\odot$}}}
\newcommand{\lta}{\raisebox{-0.6ex}{$\,\stackrel
{\raisebox{-.2ex}{$\textstyle <$}}{\sim}\,$}}

\newcommand{\specsfig}[1]        % Double FIGures (put two figures  
{
   \begin{center}
     \begin{minipage}[t]{0.45\textwidth}
         \psfig{file=#1.eps,height=0.9\textwidth,angle=270}
     \end{minipage}
     \end{center}
 }

\newcommand{\specdfig}[2]        % Double FIGures (put two figures  
{
   \begin{center}
     \begin{minipage}[t]{0.45\textwidth}
         \psfig{file=#1.eps,height=0.9\textwidth,angle=270}
     \end{minipage}
     \hfill
     \begin{minipage}[t]{0.45\textwidth}
         \psfig{file=#2.eps,height=0.9\textwidth,angle=270}
     \end{minipage}
   \end{center}
}

\begin{document}

\title[High-resolution 37.7-GHz methanol masers]{The first high-resolution observations of 37.7-, 38.3- and 38.5-GHz methanol masers}
\author[Ellingsen \etal]{S.\ P. Ellingsen,$^{1}$\thanks{Email: Simon.Ellingsen@utas.edu.au} M.\ A. Voronkov,$^{2}$ S.\ L. Breen,$^{3,2}$ J.\ L. Caswell,$^{2}$ A.\ M. Sobolev$^{4}$\\
  \\
  $^1$ School of Natural Sciences, University of Tasmania, Private Bag 37, Hobart, Tasmania 7001, Australia\\
  $^2$ CSIRO Astronomy and Space Science, PO Box 76, Epping, NSW 1710, Australia\\
  $^3$ Sydney Institute for Astronomy (SIfA), School of Physics, University of Sydney, Sydney, NSW 2006, Australia\\
  $^4$ Astronomical Observatory, Ural Federal University, Lenin avenue 51, 620000 Ekaterinburg, Russia}

 \maketitle
 
\begin{abstract}
We have used the Australia Telescope Compact Array (ATCA) to undertake the first high angular resolution observations of 37.7-GHz ($7_{-2} \rightarrow 8_{-1}E$) methanol masers towards a sample of eleven high-mass star formation regions which host strong 6.7-GHz methanol masers.  The 37.7-GHz methanol sites are coincident to within the astrometric uncertainty (0.4 arcseconds) with the 6.7-GHz methanol masers associated with the same star formation region.  However, spatial and spectral comparison of the 6.7- and 37.7-GHz maser emission within individual sources shows that the 37.7-GHz masers are less often, or to a lesser degree co-spatial than are the 12.2-GHz and 6.7-GHz masers.  We also made sensitive, high angular resolution observations of the 38.3- and 38.5-GHz class II methanol transitions ($6_{2} \rightarrow 5_{3}\mbox{A}^-$ and $6_{2} \rightarrow 5_{3}\mbox{A}^+$, respectively) and the 36.2-GHz ($4_{-1} \rightarrow 3_{0}E$) class I methanol transition towards the same sample of eleven sources.  The 37.7-, 38.3- and 38.5-GHz methanol masers are unresolved in the current observations, which implies a lower limit on the brightness temperature of the strongest masers of more than $10^6$~K.   We detected the 38.3-GHz methanol transition towards 7 sources, 5 of which are new detections and detected the 38.5-GHz transition towards 6 sources, 4 of which are new detections.  We detected 36.2-GHz class~I methanol masers towards all eleven sources, 6 of these are new detections for this transition, of which 4 sources do not have previously reported class~I methanol masers from any transition.  
\end{abstract}

\begin{keywords}
masers --- stars:formation --- ISM: molecules
\end{keywords}

\section{Introduction}

The details of the physical processes that govern the earliest stages of high-mass star formation remain an area of intense research and much debate \citep[e.g.][]{Tan+14}.  Class~II methanol masers are widely recognised as one of the most reliable tracers of the earliest stages of high-mass star formation \citep{Ellingsen06} and have the advantage that they are exclusively associated with this phenomenon \citep{Breen+13b}.  

The rich microwave spectrum of the methanol molecule contributes to the large number of different transitions which show maser emission.  Methanol masers are empirically divided into two classes, known as class~I and class~II \citep{Menten91b}.  The population inversion for class~I methanol masers is due to collisions\citep{Cragg+92,Sobolev+07,Leurini+16}, with the 44- and 36-GHz transitions the most commonly observed and generally strongest transitions \citep[e.g.][]{Voronkov+14}.  The class~I methanol masers are associated with molecular outflows \citep[e.g.][]{Plambeck+90,Voronkov+06} and other phenomenon which produce low-velocity shocks in molecular gas, such as expansion of \ionhy regions \citep[e.g.][]{Voronkov+10a} and cloud-cloud collisions \citep[e.g.][]{Sobolev92,Sjouwerman+10b}.   Searches for class~I methanol masers were initially limited by insufficient target sites known to host outflows.  However, searches towards class~II methanol maser sites \citep[e.g.][]{Slysh+94,Valtts+99,Ellingsen05} and more recently sources showing excess 4.5-\micron\/ emission in {\em Spitzer} images \citep[known as extended green objects - EGOs][]{Cyganowski+08}, have been very successful \citep[e.g.][]{Chen+11,Chen+13a} and there are now more than 500 class~I methanol maser sites known throughout the Milky Way \citep{Yang+17}.  The limited sensitive and unbiased searches for class~I methanol masers undertaken to date \citep[e.g.][]{Jordan+15,Jordan+17,Yusef-Zadeh+13}, suggest that the ATCA Legacy survey of the Galactic Plane at 7mm which commenced in 2017 will more than double the number of known sources over the next few years.

Infrared radiation creates the population inversion in class~II methanol masers \cite[e.g.][]{Sobolev+97a,Cragg+05}, with the 6.7- and 12.2-GHz transitions the most commonly observed and generally strongest.  The majority of studies to date have focused on the class~II methanol masers, as the strong transitions are at lower frequencies, (hence more readily observed) and early targeted searches towards other maser transitions and infrared sources were very successful \citep[e.g.][]{Menten91a,MacLeod+92,Caswell+95a,Walsh+98}.  There are now more than 1000 class~II methanol maser sites known in the Milky Way, with a number of large-scale, complete searches having been undertaken \citep[e.g.][]{Ellingsen+96b,Pandian+07}, of which the Methanol Multibeam (MMB) is the most comprehensive \citep{Green+10,Caswell+10,Caswell+11,Green+12a,Breen+15}.  

A total of 18 different class~II methanol maser transitions or transition series have been observed toward high-mass star formation regions \citep[see][and references therein]{Ellingsen+12b}.  The majority of these transitions have only been detected in a small number of sources and have significantly weaker emission than is observed from the 6.7-GHz transition.  Targeted searches have been made at 12.2-GHz towards all 6.7-GHz methanol masers detected in the MMB \citep{Breen+12a,Breen+12b,Breen+14a,Breen+16}, with detections towards 45.3 per cent of the sources.  However, the only other class~II methanol maser transitions for which there have been sensitive searches towards moderately large  samples of sources are the 19.9-GHz transition \citep[][22 sources]{Ellingsen+04}, the 23.1-GHz transition \citep[][50 sources]{Cragg+04}, the 37.7-, 38.3- and 38.5-GHz transitions \citep[][106 sources]{Haschick+89,Ellingsen+11a,Ellingsen+13a}, the 85.5-, 86.6- and 86.9-GHz transitions \citep[][22 sources]{Ellingsen+03} and the 107- and 156.6-GHz transitions \citep[][80 sources]{Caswell+00}.  Where multiple class~II methanol maser transitions are observed to be co-spatial there is the potential to use them as probes of the physical conditions within the star formation regions \citep[e.g.][]{Cragg+01,Sutton+01}.  However, for many of the maser transitions observations have only been made with single-dish telescopes with angular resolutions of order arcminutes.  Interferometric observations are required to determine whether the different transitions are truly co-spatial, and hence whether the underlying assumption of the multi-transition maser modelling is valid \citep[e.g.][]{Krishnan+13}.  Here we present the first interferometric observations of the 37.7-, 38.3- and 38.5-GHz class~II methanol maser transitions.  All of the sources in our sample have interferometric observations at 6.7-GHz available in the published literature, and many have also been observed at arcsecond angular resolution in the 12.2- and 19.9-GHz transitions, enabling us to test the degree to which the emission is co-spatial over these transitions.

\section{Observations}
The observations were undertaken with the Australia Telescope Compact Array (ATCA) radio telescope on 2011 March 24 (project code C2191).  The target sources were the eleven 37.7-GHz methanol masers from \citet{Ellingsen+11a} within the declination range of the ATCA.  The array was in the EW367 configuration and the synthesised beam width for the observations at 38~GHz was approximately 3 arcseconds $\times$ 5 arcseconds (although it was significantly more elongated in declination for two near-equatorial sources).  The EW367 array has baseline lengths for the inner five antennas ranging from 46 to 367 metres and the baselines from those antennas to sixth antenna are 4.0 -- 4.4~km.  In general only the short baselines are used from observations made in this array configuration and this is the case for the majority of the investigations undertaken here, although we do use it to place limits on the brightness temperature. The astrometric accuracy of ATCA observations is primarily determined by the atmospheric conditions and the properties of the phase calibrators.  The weather during the observations was excellent with the array seeing monitor reporting RMS path length noise of $\sim$ 70~$\mu$m and the astrometric accuracy is estimated to be approximately 0.4 arcseconds (see Section~\ref{sec:res37}).

The Compact Array Broadband Backend (CABB) was used for the observations \citep{Wilson+11} and configured with two 2 GHz bandwidths, each with a total of 32 $\times$ 64 MHz channels.  At the time of the observations a single 64 MHz zoom band with 2048 spectral channels was available for each 2 GHz IF.  For one of the IFs we alternated the zoom band (and IF) centre frequency between the 37710~MHz and 36175~MHz to cover the 37.7-GHz class II methanol maser transition and the 36.2-GHz class I methanol maser transition, respectively.  For the other IF we alternated the zoom band (and IF) centre frequencies between 38300~MHz and 38460~MHz to cover the 38.3- and 38.5-GHz class II methanol maser transitions, respectively.  Each zoom band has a total velocity coverage of $\sim$ 500~\kms\/ and a velocity resolution of 0.3~\kms\/ for a uniform weighting of the correlation function.

The data were reduced with {\sc miriad} using the standard techniques for ATCA spectral line observations, including correction for atmospheric opacity.  Amplitude calibration was with respect to Uranus and  PKS\,B1253$-$055 was observed as the bandpass calibrator.  In 2016 the ATCA updated the flux density model for PKS\,B1934$-$638 at higher frequencies and it became the recommended flux density calibrator for 7-mm observations.  We also observed PKS\,B1934$-$638 during our observations and reprocessed the data for several of the 37.7-GHz methanol masers using PKS\,B1934$-$638 as the flux density calibrator and found agreement to within a few percent with the values obtained using Uranus.  The absolute flux density calibration is estimated to be accurate to 10 percent. The observing strategy interleaved 3 minutes onsource for two or three maser targets with 2 minute observations of a phase calibrator before and after the target source block.  The pointing centres for the observation of each source are summarised in Table~\ref{tab:observations}, along with information on the total time onsource for each source and for each transition and the phase calibrator used.  The resulting RMS noise in a single 0.25~\kms\/ spectral channel image was $\lta$ 25 mJy beam$^{-1}$.  We adopted rest frequencies (uncertainty), of 36.169290(0.000014), 37.703696(0.000013), 38.293292(0.000014) and 38.452653(0.000014) GHz for the $4_{-1} \rightarrow 3_{0}E$, $7_{-2} \rightarrow 8_{-1}E$, $6_{2} \rightarrow 5_{3}A^-$ and $6_{2} \rightarrow 5_{3}A^+$ transitions respectively \citep{Xu+97}.  The uncertainties in the rest frequencies are all similar and correspond to an uncertainty in velocity of approximately 0.11~\kms. 

For each transition where emission was detected the absolute position was measured by imaging a spectral channel (usually of width 0.25~\kms\/, except for the weakest sources) and fitting the synthesised beam profile to the image data.  For sources with a peak flux density stronger than approximately 300~mJy we then undertook a single iteration of phase self-calibration using a model created from the clean components of the initial imaging.  In all cases this improved the signal-to-noise ratio in the image cubes, although in most cases the improvement was modest, consistent with good a priori calibration and the excellent weather conditions under which the observations were made.  To image any radio continuum emission in each field we combined the line-free data from each of the four zoom bands.  To increase the sensitivity of the continuum images to large-scale structure we used {\sc miriad}'s multi-frequency synthesis option in the {\tt invert} task and set Brigg's robustness parameter to 1.  We have not made any primary beam corrections in the current observations, as the majority of the emission from all transitions is in the central region of the primary beam where this is not necessary.  Furthermore, the accuracy of the primary beam corrections depends critically on having a good model of the beam, which has recently been found to not be the case for the ATCA at 7-mm.

\begin{table*}
\caption{The coordinates of the pointing centre for each of the class~II methanol maser sites observed.  The RMS values refer to images made without any self-calibration.}
 \begin{tabular}{lllccccc} \hline
      \multicolumn{1}{c}{\bf Source} & \multicolumn{1}{c}{\bf RA}  & \multicolumn{1}{c}{\bf Dec} & \multicolumn{1}{c}{\bf Phase} & \multicolumn{2}{c}{\bf Time on source (min)} & \multicolumn{2}{c}{\bf RMS in 0.25~\kms\/ channel}  \\
      \multicolumn{1}{c}{\bf name}     & \multicolumn{1}{c}{\bf (J2000)} & \multicolumn{1}{c}{\bf (J2000)} &  \multicolumn{1}{c}{\bf calibrator} & \multicolumn{1}{c}{\bf 37.7/38.3~GHz} & \multicolumn{1}{c}{\bf 36.2/38.5~GHz} & \multicolumn{1}{c}{\bf 37.7/38.3~GHz} & \multicolumn{1}{c}{\bf 36.2/38.5~GHz} \\  
  & \multicolumn{1}{c}{\bf $h$~~~$m$~~~$s$}& \multicolumn{1}{c}{\bf $^\circ$~~~$\prime$~~~$\prime\prime$} & & & & \multicolumn{1}{c}{\bf (mJy)} & \multicolumn{1}{c}{\bf (mJy)}\\ \hline \hline   
 G\,318.948$-$0.196 & 15 00 55.4 & $-$58 58 53 & 1613$-$586 & 18 & 18 & 23/23 & 21/23 \\ 
 G\,323.740$-$0.263 & 15 31 45.6 & $-$56 30 50 & 1613$-$586 & 18 & 18 & 24/23 & 23/23 \\ 
 G\,337.705$-$0.053 & 16 38 29.7 & $-$47 00 35 & 1646$-$50  & 18 & 18 & 24/23 & 23/23 \\ 
 G\,339.884$-$1.259 & 16 52 04.8 & $-$46 08 34 & 1646$-$50  & 18 & 18 & 22/22 & 23/21 \\ 
 G\,340.785$-$0.096 & 16 50 14.8 & $-$44 42 25 & 1646$-$50 & 18 & 18 & 24/23 & 23/23  \\ 
 G\,345.010$+$1.792 & 16 56 47.7 & $-$40 14 26 & 1646$-$50 & 15 & 12 & 23/24 & 23/26 \\ 
 G\,348.703$-$1.043 & 17 20 04.1 & $-$38 58 30 & 1759$-$39 & 15 & 12 & 23/24 & 25/25 \\ 
  NGC6334F                & 17 20 53.4 & $-$35 47 00 & 1759$-$39 & 15 & 12 & 24/24 & 27/26 \\ 
 G\,9.621$+$0.196    &18 06 14.8 & $-$20 31 32 & 1829$-$106 & 12 & 9 & 25/24 & 27/26 \\ 
  G\,23.440$-$0.182   & 18 34 39.2 & $-$08 31 24 & 1829$-$106 &  12 & 9 & 25/26 & 30/28 \\ 
 G\,35.201$-$1.736   & 19 01 45.5 & $+$01 13 29 & 1829$-$106 & 12 & 9 & 16/17 & 30/14 \\ \hline 
\end{tabular} \label{tab:observations}
\end{table*}

\section{Results}

\subsection{37.7-GHz methanol masers and comparison with Mopra spectra} \label{sec:res37}

The sample for our study were the eleven $7_{-2} \rightarrow 8_{-1}E$ (37.7-GHz) methanol maser sources south of declination +5$^\circ$ observed by \citet{Ellingsen+11a}.   Table~\ref{tab:res37} summarises the properties of the ten  37.7-GHz methanol masers detected in the current observations and the spectra extracted from the self-calibrated image cubes are shown in Figure~\ref{fig:meth37}.  The velocity ranges given in Table~\ref{tab:res37} are the range over which the image cube shows emission present at the maser location with an intensity of 4-$\sigma$ or greater ($\sim$ 100~mJy).  The spectra shown in Fig.~\ref{fig:meth37} cover the entire velocity range of the 6.7-GHz methanol masers in each of the sources. Due to the large difference between the peak intensity and the weaker emission in many sources we have included scaled versions of the spectra of stronger sources in Figure~\ref{fig:meth37} to facilitate investigation of all maser components.

The current ATCA observations detected 37.7-GHz emission from ten of the eleven sources observed, the only non-detection being G\,35.201$-$1.736, for which we are able to place an upper limit (5-$\sigma$) on the 37.7-GHz emission at the epoch of the observations of $<$80~mJy.  \citet{Ellingsen+11a} detected 37.7-GHz emission towards G\,35.201$-$1.736 with a peak flux density of 3.2~Jy in 2005 December, but observations in 2009 June and 2011 February did not detect any emission with upper limits of 1.5 and 1.1~Jy, respectively.  The current observations show that in 2011 May the 37.7-GHz methanol maser emission is at least a factor of 40 weaker than 5.5 years earlier.

There are ten 37.7-GHz methanol masers detected in the current observations that were also observed by \citet{Ellingsen+11a} and in addition the 38.3- and 38.5-GHz masers towards G\,345.010$+$1.792 and NGC\,6334F were also detected in those data.  This gives a total of 14 spectra that can be compared between the current observations and those of \citeauthor{Ellingsen+11a}.  We find that for 9 of the 14 cases the current observations have weaker emission.    Excluding the sources less than 5~Jy, for which the noise in the Mopra observations makes the amplitude measurements unreliable, the median across the sample shows the flux density measured in the current ATCA observations is 0.93 times that reported in the Mopra observations made two years previous.  This is consistent with the estimated accuracy of the Mopra and ATCA flux density scales (15 per cent and 10 per cent respectively).

Some of the sources have multiple spectral features which show differences in the relative intensity between the Mopra and ATCA observations (e.g. G\,323.740$-$0.263) and this, combined with the non-detection of G\,35.201$-$1.736 demonstrates that at least in some sources the 37.7-GHz transition can show significant temporal variability.  If we exclude the 37.7-GHz methanol masers with a peak flux density $<$ 5~Jy from consideration (as these are most susceptible to the effects of noise), the ratios between the Mopra and ATCA peak flux densities range from $\sim$0.6 -- 1.3.  So our results are consistent with modest levels of variability being present in some 37.7-GHz methanol masers on timescales of 2 years.  

Table~\ref{tab:res37} also  gives the offset between the absolute position of the 6.7-GHz methanol maser emission determined by \citet{Caswell09} and the current 37.7-GHz methanol maser observations (ten sources).  The offsets range from 0.3 -- 1.5~arcseconds, with eight of the ten being less than 1 arcsecond.  In most cases the absolute positions of the two transitions have been determined from spectral peaks with similar velocities, however, even where that is not the case, the strong 6.7-GHz methanol maser emission is generally confined to regions with an angular extent of around 0.2 -- 0.3~arcseconds \citep[e.g.][]{Phillips+98b,Walsh+98}, which will not produce a significant offset.  The astrometric accuracy of ATCA observations with high signal-to-noise and good calibration is $\sim$0.4 arcseconds \citep{Caswell97}.  The current observations employed four different phase calibrators to accommodate target sources across a range of Galactic Longitudes, we have imaged each phase calibrator using the adjacent calibrator and measured the position and find the mean offset to be 0.4 arceconds, consistent with expectations.   The mean offset between the 6.7- and 37.7-GHz methanol masers is 0.68 arcseconds, with a median of 0.58 arcseconds.  This difference is consistent with no offset between two independent observations each with an astrometric uncertainty of around 0.4 arcseconds.  Hence we conclude that to within the astrometric uncertainty of the current observations the 6.7- and 37.7-GHz methanol masers arise from the same location.

\begin{table*}
\caption{The properties of the 37.7-GHz methanol maser emission detected.  The RMS values refer to the image data from which the spectra have been extracted.}
 \begin{tabular}{lllrrr@{,}lcc} \hline
      \multicolumn{1}{c}{\bf Source} & \multicolumn{1}{c}{\bf RA}  & \multicolumn{1}{c}{\bf Dec} & \multicolumn{1}{c}{\bf Peak Flux} & \multicolumn{1}{c}{\bf Peak} & \multicolumn{2}{c}{\bf Velocity} & \multicolumn{1}{c}{\bf RMS in 0.25~\kms} & \multicolumn{1}{c}{\bf Offset from} \\
      \multicolumn{1}{c}{\bf name}     & \multicolumn{1}{c}{\bf (J2000)} & \multicolumn{1}{c}{\bf (J2000)} &  \multicolumn{1}{c}{\bf Density} & \multicolumn{1}{c}{\bf Velocity} & \multicolumn{2}{c}{\bf Range} &\multicolumn{1}{c}{\bf channel} & \multicolumn{1}{c}{\bf 6.7-GHz position}\\
  & \multicolumn{1}{c}{\bf $h$~~~$m$~~~$s$}& \multicolumn{1}{c}{\bf $^\circ$~~~$\prime$~~~$\prime\prime$} & \multicolumn{1}{c}{\bf (Jy)} & \multicolumn{1}{c}{\bf (\kms)} & \multicolumn{2}{c}{\bf (\kms)} & \multicolumn{1}{c}{\bf (mJy)}  & \multicolumn{1}{c}{\bf (arcsec)} \\   \hline \hline   
 G\,318.948$-$0.196 & 15 00 55.32 & $-$58 58 51.9 &   5.9 & $-$34.2 & $-$36.0 & $-$31.5 & 23 & 1.05 \\  
 G\,323.740$-$0.263 & 15 31 45.39 & $-$56 30 49.3 & 16.1 & $-$51.2 & $-$54.5 & $-$48.3 & 23 & 0.94 \\ 
 G\,337.705$-$0.053 & 16 38 29.65 & $-$47 00 35.9 & 2.0 & $-$55.0 & $-$56.0 & $-$52.3 & 23 & 0.45 \\  
 G\,339.884$-$1.259 & 16 52 04.70 & $-$46 08 34.6 & 280 & $-$38.7 & $-$41.5 & $-$33.0 & 23 & 0.58 \\ 
 G\,340.785$-$0.096 & 16 50 14.83 & $-$44 42 26.5 & 14.4 & $-$105.5 & $-$107.3 & $-$99.0 & 24 & 0.23 \\ 
 G\,345.010$+$1.792 & 16 56 47.57 & $-$40 14 25.5 & 181.0 & $-$22.0 & $-$25.2 & $-$11.5 & 24 & 0.32 \\  
 G\,348.703$-$1.043 & 17 20 04.04 & $-$38 58 31.1 & 2.2 & $-$3.5 & $-$3.7 & $-$3.0 & 24 & 0.31 \\ 
 NGC6334F                & 17 20 53.40 & $-$35 47 01.9 & 38.9 & $-$10.8 & $-$12.5 & $-$5.3 & 24 & 0.79 \\ 
 G\,9.621$+$0.196    &18 06 14.65 & $-$20 31 31.9 & 14.9 & $-$1.0 & $-$2.3 & $+$4.0 & 24 & 0.57 \\ 
 G\,23.440$-$0.182   & 18 34 39.19 & $-$08 31 25.8 & 1.1 & 98.0 & 96.3 & 98.8 & 24 & 1.51 \\ \hline 
\end{tabular} \label{tab:res37}
\end{table*}

\begin{figure*}
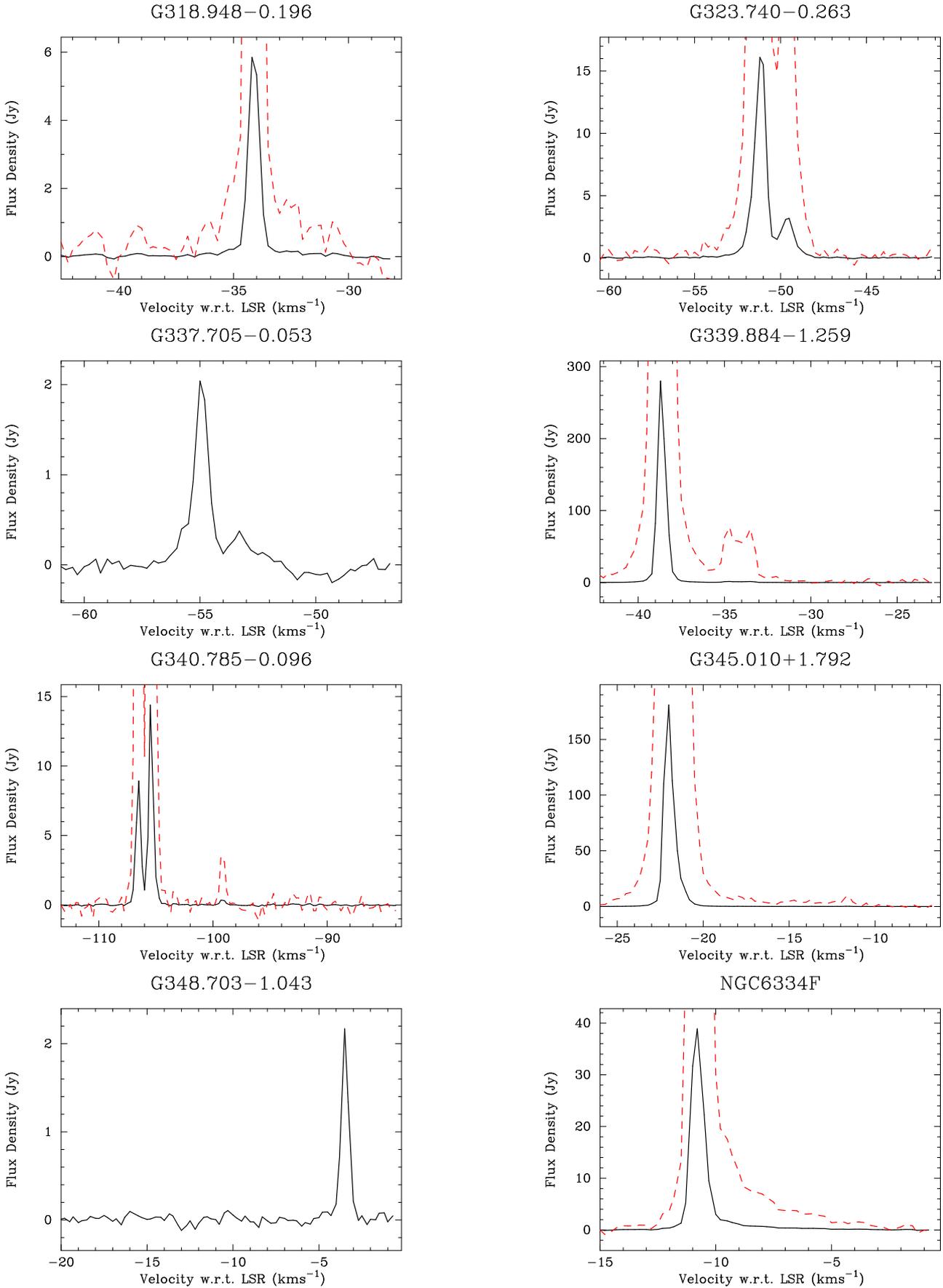

  \specdfig{g318.948_37ghz}{g323.740_37ghz}
  \specdfig{g337.705_37ghz}{g339.884_37ghz}
  \specdfig{g340.785_37ghz}{g345.010_37ghz}
  \specdfig{g348.703_37ghz}{ngc6334f_37ghz}
  \caption{Spectra of the 37.7-GHz methanol maser emission extracted from the image cubes (solid line).  The dashed line shows a scaled version of the spectrum to highlight weaker spectral features.  For sources with a peak flux density between 5 and 100~Jy the dashed line is the spectrum scaled by a factor of 10 and for those with a peak flux density greater than 100~Jy the scaling is a factor of 50.}
  \label{fig:meth37}
\end{figure*} 

\begin{figure*}
  \specdfig{g9.621_37ghz}{g23.440_37ghz}
  \contcaption{}
\end{figure*} 

\subsection{38.3- and 38.5-GHz methanol emission and absorption}

\begin{figure*}
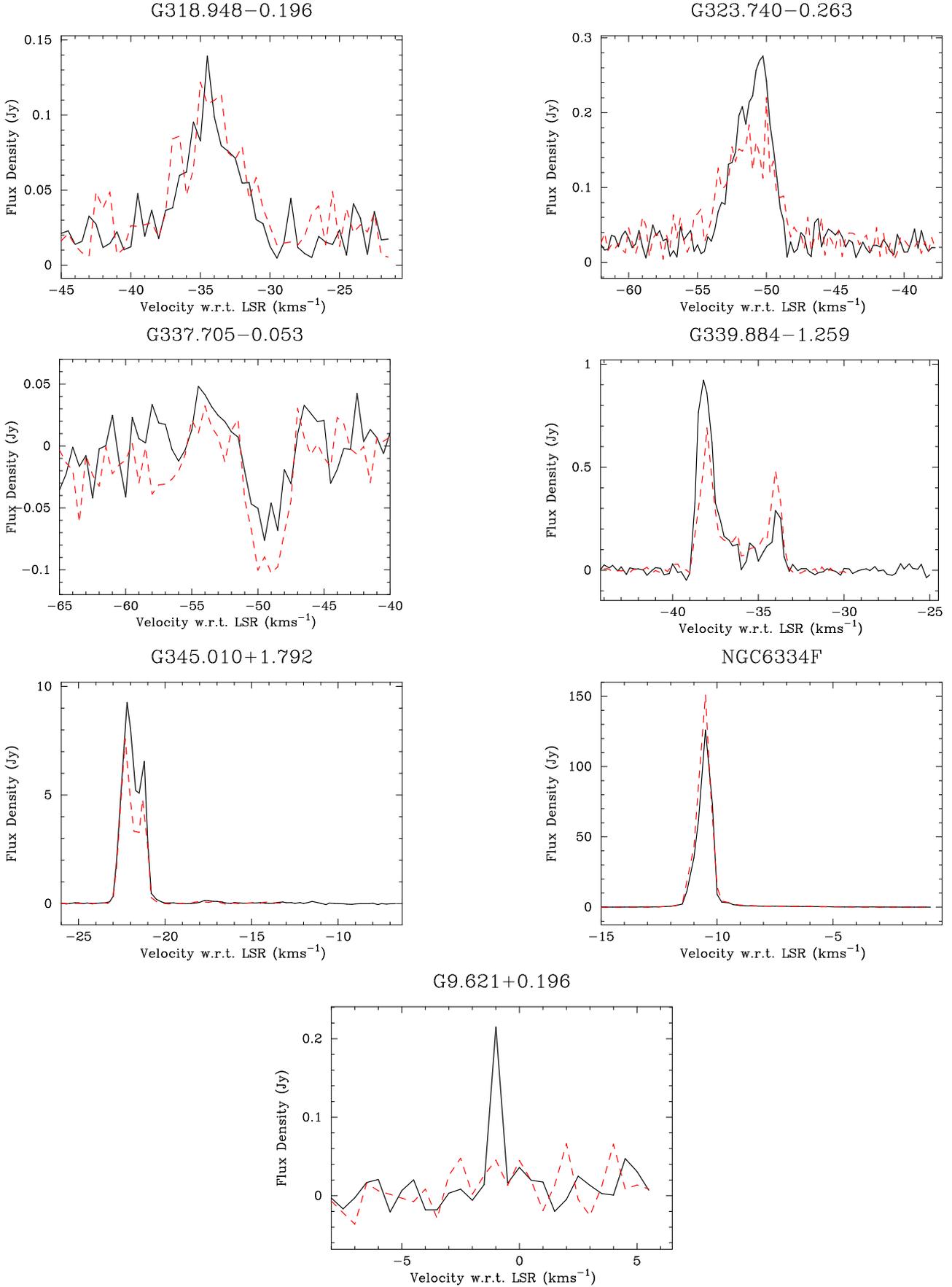

  \specdfig{g318.948_38_comp}{g323.740_38_comp}
  \specdfig{g337.705_38_comp}{g339.884_38_comp}
  \specdfig{g345.010_38_comp}{ngc6334f_38_comp}
  \specsfig{g9.621_38_comp}
  \caption{Spectra of the 38.3-GHz (solid line) and 38.5 GHz (dashed line) methanol maser emission.  The spectra were extracted from vector averaged uv data for G\,318.948$-$0.196, G\,323.740$-$0.263 and G\,337.705$-$0.053, and from the self-calibrated image cubes for the other sources.}
  \label{fig:meth38}
\end{figure*}

The 38.3-GHz methanol transition ($6_{2} \rightarrow 5_{3}A^-$) was detected towards seven of the eleven sources observed, of these two are likely not masers and in one source absorption is detected.  The 38.5-GHz methanol transition ($6_{2} \rightarrow 5_{3}A^+$) was detected towards six sources (all the sources detected in the 38.3-GHz transition with the exception of G\,9.621$+$0.196), of these two are likely not masers and in one source absorption is detected.    The properties of the 38.3- and 38.5-GHz methanol detections are summarised in Tables~\ref{tab:res38_3} and \ref{tab:res38_5} respectively, while the spectra for both transitions are shown in Figure~\ref{fig:meth38}.  For four of these sources (G\,318.948$-$0.196, G\,323.740$-$0.263, G\,339.884$-$1.259 and G\,9.621$+$0.196)  these observations represent the first reported detection of emission in either the 38.3- or 38.5-GHz transitions.  For all these new detections the peak flux density of the emission is $<$ 1~Jy and so not detectable in previous single dish observations with lower sensitivity.  The 38.3- and 38.5-GHz emission towards G\,318.948$-$0.196 and G\,323.740$-$0.263 has a weak and broad spectral profile and is likely to be thermal, or quasi-thermal emission, rather than a maser.    While for G\,337.705$-$0.053, we detect weak absorption in both the 38.3- and 38.5-GHz transitions.  For G\,339.884$-$1.259 the emission shows a much more typical class~II methanol maser profile and is stronger, with peak flux densities around 0.9~Jy, while for G\,9.621$+$0.196, we see a single narrow peak in the $6_{2} \rightarrow 5_{3}A^-$ transition, but no significant emission detected in the $6_{2} \rightarrow 5_{3}A^+$ transition.

\begin{table*}
\caption{The properties of the 38.3-GHz methanol transition observations.  The sources where the names are given in italics are new detections of this transition from the current observations.  The spectral channels used to measure the RMS were 0.25~\kms\/ for most sources, but 0.5~\kms\/ for those indicated with $^*$ and 1.0~\kms\/ for those indicated with $^\dagger$.  The RMS values refer to the image data from which the spectra have been extracted.  NOTE: For G\,337.705$-$0.053 the absorption in this transition was too weak to reliably image, so the position is assumed to be the same as for the 38.5-GHz transition.}
 \begin{tabular}{lllrrr@{,}lcc} \hline
      \multicolumn{1}{c}{\bf Source} & \multicolumn{1}{c}{\bf RA}  & \multicolumn{1}{c}{\bf Dec} & \multicolumn{1}{c}{\bf Peak Flux} & \multicolumn{1}{c}{\bf Peak} & \multicolumn{2}{c}{\bf Velocity} & \multicolumn{1}{c}{\bf RMS in spectral} & \multicolumn{1}{c}{\bf Offset from} \\
      \multicolumn{1}{c}{\bf name}     & \multicolumn{1}{c}{\bf (J2000)} & \multicolumn{1}{c}{\bf (J2000)} &  \multicolumn{1}{c}{\bf Density} & \multicolumn{1}{c}{\bf Velocity} & \multicolumn{2}{c}{\bf Range} &\multicolumn{1}{c}{\bf channel} & \multicolumn{1}{c}{\bf 37.7-GHz position}\\
  & \multicolumn{1}{c}{\bf $h$~~~$m$~~~$s$}& \multicolumn{1}{c}{\bf $^\circ$~~~$\prime$~~~$\prime\prime$} & \multicolumn{1}{c}{\bf (Jy)} & \multicolumn{1}{c}{\bf (\kms)} & \multicolumn{2}{c}{\bf (\kms)} & \multicolumn{1}{c}{\bf (mJy)}  & \multicolumn{1}{c}{\bf (arcsec)} \\   \hline \hline   
 {\em G\,318.948$-$0.196} & 15 00 55.35 & $-$58 58 52.2 & 0.14 & $-$34.5 & $-$36 & $-$32 & 8$^\dagger$ & 0.38 \\  
 {\em G\,323.740$-$0.263} & 15 31 45.39 & $-$56 30 49.8 & 0.28 & $-$50.2 & $-$54 & $-$48.5 & 8$^*$ & 0.50\\ 
 {\em G\,337.705$-$0.053} & 16 38 29.70 & $-$47 00 35.1 & $-$0.08 & $-$49.5 & $-$51 & $-$47 & 5$^*$ & 0.95 \\
 {\em G\,339.884$-$1.259} & 16 52 04.71 & $-$46 08 34.8 & 0.93 & $-$38.3 & $-$39 & $-$33 & 15 & 0.22 \\ 
 G\,345.010$+$1.792 & 16 56 47.59 & $-$40 14 25.6 & 9.3 & $-$22.3 & $-$24 & $-$16.5 & 13 & 0.35 \\  
 NGC6334F                & 17 20 53.37 & $-$35 47 02.0 & 126 & $-$10.5 & $-$13 & $-$3 & 12 & 0.38 \\ 
 {\em G\,9.621$+$0.196}    &18 06 14.60 & $-$20 31 31.4 & 0.22 & $-$1.0 & $-$1.5 & $-$0.5 & 8$^*$ & 0.86 \\ \hline 
\end{tabular} \label{tab:res38_3}
\end{table*}

\begin{table*}
\caption{The properties of the 38.5-GHz methanol maser transition observations.  The sources where the names are given in italics are new detections of this transition from the current observations.  The spectral channels used to measure the RMS were 0.25~\kms\/ for most sources, but 0.5~\kms\/ for those indicated with $^*$ and 1.0~\kms\/ for those indicated with $^\dagger$.  The RMS values refer to the image data from which the spectra have been extracted}
 \begin{tabular}{lllrrr@{,}lcc} \hline
      \multicolumn{1}{c}{\bf Source} & \multicolumn{1}{c}{\bf RA}  & \multicolumn{1}{c}{\bf Dec} & \multicolumn{1}{c}{\bf Peak Flux} & \multicolumn{1}{c}{\bf Peak} & \multicolumn{2}{c}{\bf Velocity} & \multicolumn{1}{c}{\bf RMS in spectral} & \multicolumn{1}{c}{\bf Offset from} \\
      \multicolumn{1}{c}{\bf name}     & \multicolumn{1}{c}{\bf (J2000)} & \multicolumn{1}{c}{\bf (J2000)} &  \multicolumn{1}{c}{\bf Density} & \multicolumn{1}{c}{\bf Velocity} & \multicolumn{2}{c}{\bf Range} &\multicolumn{1}{c}{\bf channel} & \multicolumn{1}{c}{\bf 37.7-GHz position}\\
  & \multicolumn{1}{c}{\bf $h$~~~$m$~~~$s$}& \multicolumn{1}{c}{\bf $^\circ$~~~$\prime$~~~$\prime\prime$} & \multicolumn{1}{c}{\bf (Jy)} & \multicolumn{1}{c}{\bf (\kms)} & \multicolumn{2}{c}{\bf (\kms)} & \multicolumn{1}{c}{\bf (mJy)}  & \multicolumn{1}{c}{\bf (arcsec)} \\   \hline \hline   
 {\em G\,318.948$-$0.196} & 15 00 55.29 & $-$58 58 52.1 & 0.12 & $-35.0$ & $-37$ & $-30.5$ & 9$^\dagger$ & 0.31 \\  
 {\em G\,323.740$-$0.263} & 15 31 45.41 & $-$56 30 49.3 & 0.22 & $-50.0$ & $-54$ & $-48.5$ & 9$^*$ & 0.17 \\ 
 {\em G\,337.705$-$0.053} & 16 38 29.70 & $-$47 00 35.1 & $-0.10$ & $-49.0$ & $-51$ & $-47$ & 5$^*$ & 0.95 \\
 {\em G\,339.884$-$1.259} & 16 52 04.70 & $-$46 08 35.0 & 0.69 & $-38.3$ & $-39$  & $-33.5$ & 13 & 0.40 \\ 
 G\,345.010$+$1.792 & 16 56 47.57 & $-$40 14 26.1 & 5.94 & $-22.3$ & $-23.3$ & $-16.8$ & 9 & 0.50 \\  
 NGC6334F                & 17 20 53.37 & $-$35 47 02.0 & 151 & $-10.5$ & $-13.5$ & $-3.5$ & 13 & 0.38 \\ \hline 
\end{tabular} \label{tab:res38_5}
\end{table*}

\subsection{36.2-GHz class I methanol masers}

The 36.2-GHz class I methanol maser transition was included in the observations because it could be observed simultaneously
with the 38.5-GHz transition with no negative impact on the quality of the 37.7-GHz data collected.  Somewhat surprisingly,
36.2-GHz class I methanol maser emission was detected towards all eleven sources observed.  The properties of the detected
maser emission are summarised in Table~\ref{tab:res36}.  The listed peak flux density and velocity are considered over all class~I methanol maser regions for that source, whereas the integrated values are extracted from the spectra shown in Figure~\ref{fig:meth36}, which have been extracted from the image cube over a region which encompasses all class~I methanol maser regions within the primary beam of each source.  A search of the published literature shows that four of the sources have not previously been reported to have class~I methanol maser emission in any transition (G\,337.705$-$0.053, G\,340.785$-$0.096, G\,348.703$-$1.043 \& G\,35.201$-$1.736).  Five of the sources (G\,318.948$-$0.196, G\,323.740$-$0.263, G\,339.884$-$1.259, G\,345.010$+$1.792 \& NGC6334F) have been previously observed at 36.2-GHz in the large survey of \citet{Voronkov+14}.  The remaining two sources (G\,9.621$+$0.196 \& G\,23.440$-$0.182) have previous detections of both 44- and 95-GHz class~I methanol masers \citep{Kurtz+04,Gan+13,Slysh+94,Valtts+00}, but not for the 36.2-GHz transition.

Succinctly presenting the results of interferometric observations of class~I methanol masers is more challenging than for the class~II methanol masers because the emission is much more distributed.  The distribution of the 36.2-GHz class~I methanol masers with respect to class~II masers, radio continuum and infrared emission in each source are shown in the figures in Appendix~\ref{sec:environ}.  The spectra of the main 36.2-GHz class~I methanol sites in each source are shown in these figures and where the sources have previously been observed by \citet{Voronkov+14}, we have used the same letter designations for the components.  The locations at which the spectra have been extracted for each 36.2-GHz methanol maser site are given in Table~\ref{tab:components}.  Figure~\ref{fig:meth36} shows the spectra extracted from the image cubes of the 36.2-GHz emission, with the exception of G\,23.440$-$0.182 for which the spectrum is from the scalar-averaged {\em uv}-data.  G\,23.440$-$0.182 shows relatively strong 36.2-GHz emission at a number of different sites with significant overlap in the velocity range of those sites and this causes artefacts in the vector averaged spectra shown in Figure~\ref{fig:g23}.  To avoid this in the spectrum shown in Figure~\ref{fig:meth36} we have used the scalar-averaged spectrum from the {\em uv}-data for G\,23.440$-$0.182.  The spectra in Fig.~\ref{fig:meth36} incorporate emission from all the 36.2-GHz methanol maser sites in each source, with the exception of G\,337.705$-$0.053 and NGC6334F, where there is one class~I methanol maser site (labelled E and F respectively) that is offset significantly from the rest and these have been excluded from the spectra.  These spectra should be similar to those obtained by single dish observations towards the same regions.  For velocities where the range of two or more individual 36.2-GHz methanol maser sites within one source overlap the flux densities in Fig.~\ref{fig:meth36} will exceed those shown for individual sites in Appendix~\ref{sec:environ}.  Table~\ref{tab:res36} summarises properties of the spectra shown in Fig.~\ref{fig:meth36}.

For the five sources previously observed with the ATCA in the 36.2-GHz transition, the flux densities shown in Figures~\ref{fig:g318} -- \ref{fig:g35} are systematically lower for the current observations than in the corresponding figure in \citet{Voronkov+14}.  This is thought to be due to a combination of coarser spectral resolution in the current observations, an update to the flux density scale at 7-mm for the ATCA since the observations of \citeauthor{Voronkov+14} and differences in the way the spectra have been extracted from the image data - extraction at a single point in the current observations, compared with extraction over a region by \citet{Voronkov+14}.

\begin{table*}
\caption{The properties of the 36.2-GHz class~I methanol maser emission detected.  The sources where the names are given in italics are new detections of this transition from the current observations.  The references for the distances are 1 = \citet{Reid+16} ; 
2 = \citet{Krishnan+15} ; 3 = \citet{Wu+14} ; 4 = \citet{Sanna+09} ; 5 = \citet{Brunthaler+09} ; 6 = \citet{Zhang+09}}
 \begin{tabular}{lllrrrrccr} \hline
      \multicolumn{1}{c}{\bf Source} & \multicolumn{1}{c}{\bf RA}  & \multicolumn{1}{c}{\bf Dec} & \multicolumn{2}{c}{\bf Peak} & \multicolumn{2}{c}{\bf Integrated}  & \multicolumn{1}{c}{\bf Velocity} & \multicolumn{1}{c}{\bf RMS in} 
      &\\
      \multicolumn{1}{c}{\bf name}     & \multicolumn{1}{c}{\bf (J2000)} & \multicolumn{1}{c}{\bf (J2000)} &  \multicolumn{1}{c}{\bf Flux} & \multicolumn{1}{c}{\bf Velocity} &  \multicolumn{1}{c}{\bf Peak Flux} & \multicolumn{1}{c}{\bf Peak Vel.} & \multicolumn{1}{c}{\bf Range} &\multicolumn{1}{c}{\bf 0.25~\kms}
      &  \multicolumn{1}{c}{\bf Distance}  \\
  & \multicolumn{1}{c}{\bf $h$~~~$m$~~~$s$}& \multicolumn{1}{c}{\bf $^\circ$~~~$\prime$~~~$\prime\prime$} & \multicolumn{1}{c}{\bf (Jy)} & \multicolumn{1}{c}{\bf (\kms)} & \multicolumn{1}{c}{\bf (Jy)} & \multicolumn{1}{c}{\bf (\kms)} & \multicolumn{1}{c}{\bf (\kms)} & \multicolumn{1}{c}{\bf channel} &  \multicolumn{1}{c}{\bf (kpc)}  \\   \hline \hline   
 G\,318.948$-$0.196 & 15 00 55.32 & $-$58 58 56.7 &   2.9 & -36.0 & 3.6 & -36.0 & -37.2 -- -26.7 & 23 & 2.4$^{1}$ \\ 
 G\,323.740$-$0.263 & 15 31 46.09 & $-$56 30 57.6 &   0.7 & -50.0 & 1.1 & -49.7 & -53.3 -- -47.2 & 23 & 3.1$^{1}$\\ 
 {\em G\,337.705$-$0.053} & 16 38 29.15 & $-$47 00 29.5 &   3.6 & -46.0 & 5.9 & -46.0 & -56.7 -- -38.0 & 23 & 3.3$^{1}$ \\  
 G\,339.884$-$1.259 & 16 52 05.48 & $-$46 08 40.9 &   1.3 & -31.8 & 1.8 & -31.7 & -33.2 -- -29.7 & 23 & 2.1$^{2}$ \\ 
 {\em G\,340.785$-$0.096} & 16 50 15.21 & $-$44 42 34.1 &   0.7 & -100.3 & 2.7 & -100.5 & -149.0 -- -92.5 & 23  & 6.2$^{1}$ \\ 
 G\,345.010$+$1.792 & 16 56 46.06 & $-$40 14 19.2 & 11.7 & -13.0  & 16.2 & -13.0 & -15.7 -- -12.0 & 26 & 1.8$^{1}$ \\  
 {\em G\,348.703$-$1.043} & 17 20 03.03 & $-$38 58 29.1 &   1.8  & -13.0 & 2.8 & -13.3 & -15.5 -- -11.5 & 25 & 1.4$^{1}$ \\  
 NGC6334F                & 17 20 49.79 & $-$35 46 49.0 &  8.9 & -8.0  & 12.6 & -7.8 & -11.5 -- -6.8  & 27 & 1.3$^{3}$ \\ 
 {\em G\,9.621$+$0.196}    &18 06 15.29 & $-$20 31 39.0   &  1.1  & 3.8 & 2.0 & 3.8 & -3.0 -- 7.5 & 26 & 5.2$^{4}$ \\   {\em G\,23.440$-$0.182}   & 18 34 38.44 & $-$08 31 27.8 & 25.4  & 102.8 & 29.5 & 102.8 & 96.0 -- 104.0 & 26 & 5.9$^{5}$ \\ 
 {\em G\,35.201$-$1.736}   & 19:01:42.43 & $+$01:13:43.3 & 1.2   &  42.0  & 1.3 & 42.0 & 42.0 -- 46.3 & 26 & 3.3$^{6}$\\ \hline  
\end{tabular} \label{tab:res36}
\end{table*}

\begin{figure*}
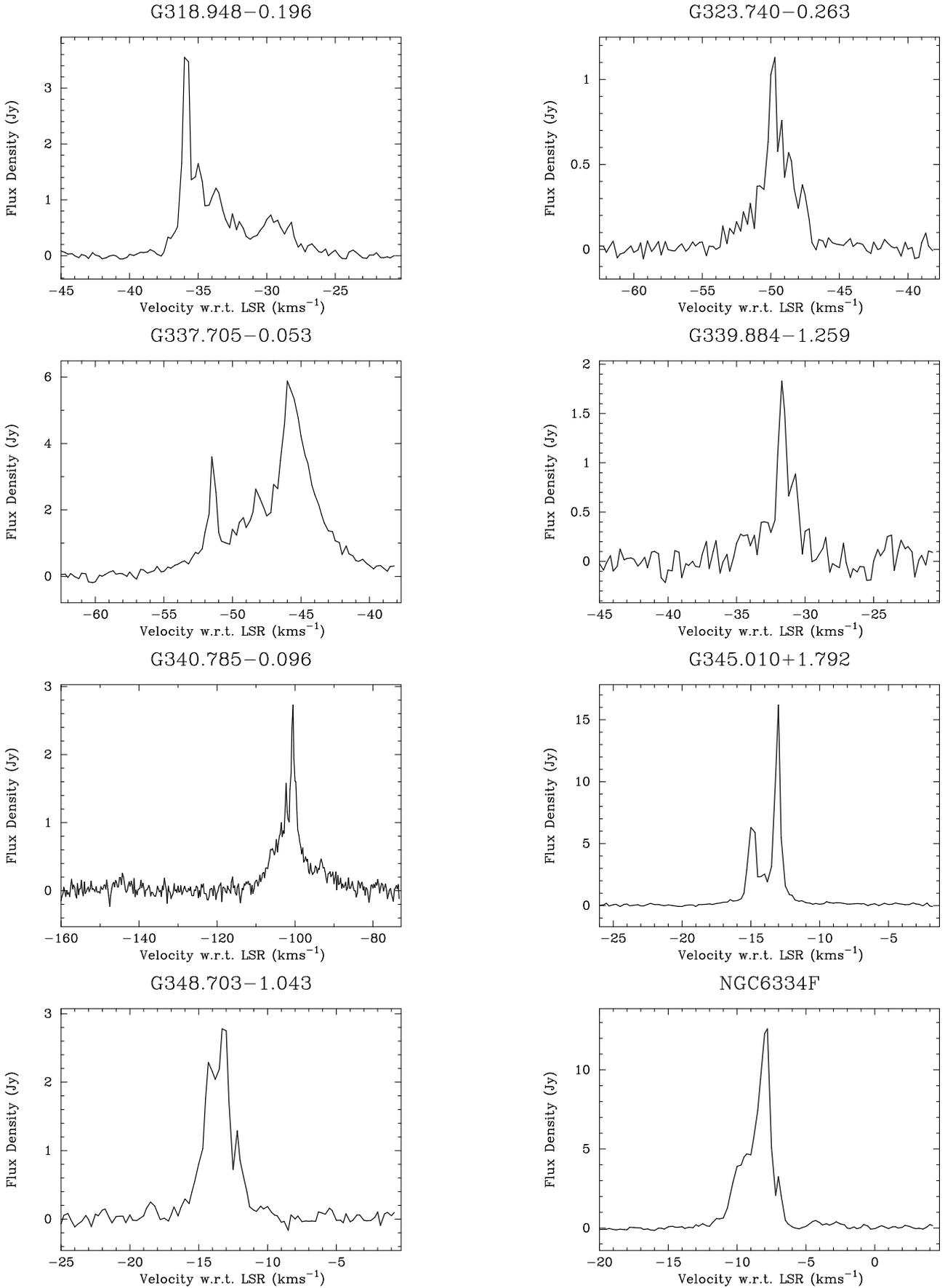

  \specdfig{g318.948_36ghz}{g323.740_36ghz}
  \specdfig{g337.705_36ghz}{g339.884_36ghz}
  \specdfig{g340.785_36ghz}{g345.010_36ghz}
  \specdfig{g348.703_36ghz}{ngc6334f_36ghz}  
  \caption{Spectra of the 36.2-GHz class~I methanol maser emission extracted from the image cube data, with the exception of G\,23.440$-$0.182, for which the spectrum is from scalar averaged {\em uv-}data.}
  \label{fig:meth36}
\end{figure*} 

\begin{figure*}
  \specdfig{g9.621_36ghz}{g23.440_36ghz_uv}
  \specsfig{g35.201_36ghz}
  \contcaption{}
\end{figure*}

\subsection{Radio continuum}

We combined the line-free data from all the zoom band observations to create a continuum dataset for each target source (effective total bandwidth of approximately 220~MHz).  We detected continuum emission toward 10 of the 11 fields, the exception was G\,$318.948-0.196$, for which we can set an upper limit on any 8-mm continuum emission of 2~mJy beam$^{-1}$ (the RMS noise level in the image is 0.24 mJy beam$^{-1}$).  The continuum emission detected toward each source is shown in Figures~\ref{fig:g323} -- \ref{fig:g35} (in black or white contours, depending on the background)  and the properties of the continuum emission are summarised in Table~\ref{tab:cont}.  For eight of the ten sources where continuum emission is observed the class~II methanol masers are projected against the continuum emission, in most cases close to the peak.  Table~\ref{tab:cont} gives the offset of the 37.7-GHz methanol maser (or 6.7-GHz methanol maser for G\,35.201$-$1.736) from the peak of the radio continuum emission.  The two sources where continuum emission is detected, but it is not associated with the maser site are G\,348.703$-$1.043 and G\,35.201$-$1.736.

\begin{table*}
\caption{The properties of the 8-mm radio continuum emission detected.  The offset is between the peak of the radio continuum and the 37.7-GHz methanol maser, except for G\,35.201$-$1.736, where the offset is with respect to the 6.7-GHz methanol maser.}
 \begin{tabular}{lllrrcc} \hline
      \multicolumn{1}{c}{\bf Source} & \multicolumn{1}{c}{\bf RA}  & \multicolumn{1}{c}{\bf Dec} & \multicolumn{1}{c}{\bf Peak Flux} & \multicolumn{1}{c}{\bf Integrated} & \multicolumn{1}{c}{\bf Image} & \multicolumn{1}{c}{\bf Offset from} \\
      \multicolumn{1}{c}{\bf name}     & \multicolumn{1}{c}{\bf (J2000)} & \multicolumn{1}{c}{\bf (J2000)} &  \multicolumn{1}{c}{\bf Density} & \multicolumn{1}{c}{\bf Flux Density} &\multicolumn{1}{c}{\bf RMS} & \multicolumn{1}{c}{\bf 37.7-GHz maser}\\
  & \multicolumn{1}{c}{\bf $h$~~~$m$~~~$s$}& \multicolumn{1}{c}{\bf $^\circ$~~~$\prime$~~~$\prime\prime$} & \multicolumn{1}{c}{\bf (mJy beam$^{-1}$)} & \multicolumn{1}{c}{\bf (mJy)} &  \multicolumn{1}{c}{\bf (mJy beam$^{-1}$)}  & \multicolumn{1}{c}{\bf (arcsec)} \\   \hline \hline   
 G\,323.740$-$0.263 & 15 31 45.71 & $-$56 30 50.7 & 2.4 & 3.8 & 0.22 & 3.0 \\ 
 G\,337.705$-$0.053 & 16 38 29.66 & $-$47 00 36.2 & 315 & 338 & 1.0 & 0.36 \\  
 G\,339.884$-$1.259 & 16 52 04.70 & $-$46 08 33.8 & 7.5 & 12.2 & 0.2 & 0.80 \\
 G\,340.785$-$0.096 & 16 50 14.83 & $-$44 42 27.0 & 43 & 46 & 0.2 & 0.5 \\
 G\,345.010$+$1.792 & 16 56 47.60 & $-$40 14 26.0 & 431 & 442 & 1.5 & 0.61 \\
 G\,348.703$-$1.043 & 17 19 58.95 & $-$38 58 15.0 & 341 & 403 & 1.5 & 62 \\ 
 NGC6334F                & 17 20 53.52 & $-$35 47 02.4 & 1990 & 2610 & 10 & 1.5 \\
 G\,9.621$+$0.196    &18 06 13.99 & $-$20 31 45.7 & 46 & 395 & 0.7 &  17 \\ 
 G\,23.440$-$0.182   & 18 34 39.14 & $-$08 31 29.7 & 1.9 & 13 & 0.4 & 4.0 \\ 
 G\,35.201$-$1.736 & 19 01 46.49 & $+$01 13 21.3 & 426 & 569 & 11 & 18$^{*}$ \\  \hline 
\end{tabular} \label{tab:cont}
\end{table*}

\subsection{Comments on Individual Sources} \label{sec:indiv}

\subsubsection{G\,318.948$-$0.196}
The sensitivity of the ATCA observations shows 37.7-GHz emission covering the velocity range -36.0 -- -31.5~\kms, although there is only a single strong spectral feature peaking at -34.2~\kms.  The self-calibrated image cube shows all the emission from the spectral channels associated with the strong component arise from the same location (scatter of less than 50 milliarcseconds in each axis). The 6.7-GHz methanol masers show around 7 spectral features \citep{Norris+93}, however, none of these match the velocity of the single strong 37.7-GHz maser peak.  The flux density of the 37.7-GHz emission detected in the current observations is approximately 30 percent lower than that reported by \citet{Ellingsen+11a}.  The environment of the G\,318.948$-$0.196 star formation region is shown in Figure~\ref{fig:g318} with the strongest 36.2-GHz class~I methanol emission offset slightly to the south of the class~II masers and distributed in an elongated structure with the same position angle as the extension in the ATLASGAL dust emission.  The class~I maser distribution shows good agreement with the observations of \citet{Voronkov+14} for this source, the only difference being that we do no detect any significant emission from component F.  \citet{Voronkov+14} find this component is broad and weak and so may be predominantly quasi-thermal emission resolved out by the higher resolution of the current observations. 

\subsubsection{G\,323.740$-$0.263:}

The 37.7-GHz methanol maser emission in this source shows two main spectral components, which the self-calibrated ATCA observations show to be offset from one another by approximately 0.2 arcseconds in the north-south direction with the red-shifted components further south.  The first high resolution images of the class~II methanol masers in this source were made by \citet{Norris+88,Norris+93} at 12.2- and 6.7-GHz respectively.  However, in contrast to most of the other sources they observed, the distribution of the 6.7- and 12.2-GHz components in this source do not align well between the two transitions. The 6.7-GHz class~II methanol masers in this source were also imaged with higher sensitivity by \citet{Phillips+98b} and their components E and C have similar velocities to the 37.7-GHz masers, however, they show a larger offset and with a different orientation (nearly diametrically opposite).  \citet{Krishnan+13} imaged the weak 19.9-GHz class~II methanol maser emission in this source, detecting a single component at a peak velocity of -50.5~\kms\/ (i.e. approximately the mid-point of the two 37.7-GHz components).  

Figure~\ref{fig:g323} shows the general environment of the G\,323.740$-$0.263 star formation region, with the class~II methanol masers associated with an infrared source with a strong excess in the 4.5-$\mu$m wavelength band, characteristic of the presence of shock-excited gas.  There is a large, evolved \ionhy region to the north-west of the maser source and the strongest class~I methanol maser emission is to the south-east, although there is also emission associated with the infrared source and close to the class~II maser site.  \citet{Voronkov+14} identified four different class~I methanol maser sites in this source, all of which showed emission from the 44-GHz transition, with B and D also showing emission at 36.2-GHz.  The current observations detect 36.2-GHz methanol masers towards all four of the \citeauthor{Voronkov+14} sites, and also at a new location to the south, which we have labelled E.  The new site is beyond the half-power point of the ATCA primary beam and Figure~\ref{fig:g323} shows that it is associated with a different ATLASGAL dust clump to the other maser emission.  Hence, it is likely an unrelated nearby site of class~I methanol masers, G\,323.734$-$0.273, not directly associated with the G\,323.740$-$0.263 maser region.

Combining the line-free channels from all four zoom-bands we are able for the first time to detect the radio continuum emission associated with the class~II methanol masers (see black contours in Figure~\ref{fig:g323}).  The emission has a peak intensity of 2.4 mJy beam$^{-1}$ and is centred at right ascension (J2000) 15$^h$31$^m$45.71$^s$, Declination $-$56$^{\circ}$30$^{\prime}$50.7$^{\prime\prime}$, an offset of 3 arcseconds to the south-east of the strongest class~II methanol maser emission. Previous ATCA observations at 8.6 and 1.6~GHz found upper limits of 0.2~mJy and 0.4~mJy, respectively on the radio continuum emission at those frequencies \citep[see][]{Voronkov+14}.

\subsubsection{G\,337.705$-$0.053:} The higher sensitivity of the ATCA observations shows the 37.7-GHz methanol masers contain at least two spectral components over a velocity range -56.0 -- -52.3~\kms.  One unique aspect of this source is that we observe absorption in the 37.7-, 38.3- and 38.5-GHz transitions centred at a velocity of approximately -49~\kms.  This can be seen in the spectrum of G\,337.705$-$0.053 in Figure~\ref{fig:meth37}, but is more readily apparent in the 38.3- and 38.5-GHz data due to the absence of emission in those transitions (see Fig.~\ref{fig:meth38}).  Images of the 38.5-GHz transition shows that the absorption is located in front of the radio continuum emission associated with the class~II methanol maser emission with the strongest absorption offset approximately 1 arcsecond north of the continuum peak.

\citet{Phillips+98b} used the ATCA to image the 6.7-GHz methanol masers in this source and found that they are distributed in an elongated structure with a simple velocity gradient and are projected across the peak of the radio continuum emission.  The secondary peak at 37.7-GHz is too weak to be able to reliably determine an accurate position for it with respect to the primary peak, so we cannot make any meaningful comparison of the relative distribution of the 6.7- and 37.7-GHz masers in this source.  

Figure~\ref{fig:g337} shows the class~II methanol maser emission and radio continuum lies at the edge of an extended infrared source, but close to the peak of the ATLASGAL 870-$\mu$m dust emission.  The bulk of the 36.2-GHz class~I methanol maser emission lies along the western-edge of the extended infrared source and the strongest emission is redshifted with respect to the 37.7-GHz methanol maser emission and the absorption seen in the other transitions.

\subsubsection{G\,339.884$-$1.259:} The higher sensitivity of the ATCA observations shows that the 37.7-GHz methanol masers have a much larger velocity range than was apparent from the single dish observations of \citet{Ellingsen+11a}.  G\,339.884$-$1.259 remains the strongest 37.7-GHz methanol maser source in the Galaxy, with a peak flux density of around 300~Jy.  In addition to the strong component there are two weaker components each with a flux density of approximately 1.5~Jy with peak velocities of -34.7 and -33.5~\kms, which can be seen in the dashed spectrum for this source in Figure~\ref{fig:meth37}.  Interestingly, G\,339.884$-$1.259 has recently been discovered to also host the strongest 4.8-GHz formaldehyde maser in the Galaxy \citep{Chen+17} and demonstrates a number of spectral components which closely align in velocity with 6.7-GHz methanol maser features.

The greater sensitivity has also resulted in the detection of weak maser emission in the 38.3- and 38.5-GHz transitions in this source (Fig.~\ref{fig:meth38}).  The profile of the two transitions is similar, with two components with peak velocities of approximately -38.5 and -34.0~\kms.  The 36.2-GHz class~I methanol maser emission in this source is relatively weak (peak flux density $<$ 2~Jy).  Figure~\ref{fig:g339} shows the location of the class~I methanol masers with respect to the other masers and continuum emission in the G\,339.884$-$1.259 star formation region.  The distribution matches that observed by \citet{Voronkov+14}, but we detect additional weak 36.2-GHz emission (component C), between components A and B.  The radio continuum emission and class~II methanol masers coincide with the peak of the infrared emission seen in the {\em WISE} image and the ATLASGAL dust continuum emission shows elongation with approximately the same position angle seen in the class~II methanol maser cluster \citep[see for example][]{Dodson08} and on larger scales in the class~I methanol maser emission.

\subsubsection{G\,340.785$-$0.096:} The notable feature of the class~II methanol maser emission in this source is the large velocity range \citep[26.5~\kms\/ in the 6.7-GHz transition;][]{Caswell+11}, only slightly less than the widest observed of 28.5~\kms\/ \citep{Green+17}.  The 37.7-GHz velocity range is also large, with the more sensitive ATCA observations detecting emission in the velocity range -107.3 -- -98.8~\kms, with a weak ($\sim$0.3~Jy) peak at around -99.3~\kms\/ which was not apparent in the Mopra observations \citep{Ellingsen+11a}.  All the strong 37.7-GHz methanol maser emission is confined to a 0.1 arcsecond square region, which, along with the restricted velocity range of the strong emission means that it isn't possible to make a detailed comparison with the complex spatial distribution of the 6.7-GHz methanol masers seen in ATCA observations \citep{Norris+93,Phillips+98b}.  

There is no 38.3- or 38.5-GHz methanol emission stronger than 100~mJy, however, we do detect 36.2-GHz class~I methanol maser emission.  Figure~\ref{fig:g340} shows the environment of the G\,340.785$-$0.096 region, with the class~II methanol masers located at the peak of an ATLASGAL dust clump, which is located within a low-contrast infrared dark cloud.  The class~II methanol masers are close to the peak of the radio continuum emission which lies between two compact infrared sources.  The class~I methanol maser emission is complex with at least 6 different sites distributed around the class~II masers and radio continuum emission.  The majority of the class~I methanol maser emission is the in the same velocity range as the class~II methanol masers, however, there are weak class~I masers at site A, to the north-east of the class~II masers which show emission in the velocity range -151.5 -- -141.0.  The most blue-shifted emission from any of the other sites is at -107.3~\kms, so this component has a velocity offset of $>$ 30~\kms\/.  \citet{Voronkov+14} found that high-velocity class~I methanol maser emission is relatively rare, with the majority of cases detected as blueshifted components in the 36.2-GHz transition.  Of the sources observed by \citet{Voronkov+14}, G\,340.785$-$0.096 appears most similar to G\,309.38$-$0.13, as the high-velocity emission has a relatively small angular separation from the other sites.  The high-velocity 36.2-GHz class~I methanol emission in G\,309.38$-$0.13 has been studied in detail by \citet{Voronkov+10b} and subsequent observations have detected 44-GHz methanol masers at the same location \citep{Voronkov+14}.  Detailed investigations of regions such as this may help us gain insight into the more extreme class~I methanol maser emission seen in the Galactic Centre \citep{Yusef-Zadeh+13} and some starburst galaxies \citep{Ellingsen+14,Ellingsen+17b,McCarthy+17}.

\subsubsection{G\,345.010$+$1.792:} This is one of the best studied class~II methanol maser sources, it has a larger number of detected class~II methanol maser transitions than any other source \citep{Ellingsen+12}.  It is the second strongest 37.7-GHz methanol maser source (after G\,339.884$-$1.259), however, it has only a single spectral component stronger than 0.5~Jy.  The 38.3- and 38.5-GHz methanol masers show similar profiles to each other, with the latter being approximately 20 percent weaker.  The two spectral components seen in the 38.3- and 38.5-GHz spectra are contained within the velocity width of the single spectral component at 37.7-GHz.  All three class~II methanol maser transitions show weak ($<$ 0.5~Jy) emission in the velocity range -20 -- -17~\kms\/.  This source also has relatively strong 36.2-GHz class I methanol maser emission which peaks at -13.0~\kms\/, well outside the velocity range of the strong class~II methanol masers and largely confined to two sites.  \citet{Voronkov+14} find three sites of 36.2-GHz methanol masers in this source and a further 4 sites with only 44.1-GHz class~I methanol masers. We detect all of the 36.2-GHz sites of \citeauthor{Voronkov+14} in the current observations, however, site G lies at the half-power point of the primary beam for the current observations and is a marginal detection, which we would not have reported without the independent confirmation from the previous observations.

G\,345.010$+$1.792 lies outside the regions covered by both the ATLASGAL \citep{Schuller+09} and GLIMPSE \citep{Benjamin+03,Churchwell+09} surveys, but has been imaged in other {\em Spitzer} observations.  Figure~\ref{fig:g345} has a {\em Spitzer} three-colour image as the background (data extracted from the archive) and shows the class~II methanol masers are projected against the radio continuum peak and are associated with a strong, extended infrared source.  The class~I methanol masers are located to the north-west of the class~II site, with some of the emission projected against a region of infrared emission which has an excess in the 4.5-$\mu$m band, indicative of shock-excited gas.

\subsubsection{G\,348.703$-$1.043:} The ATCA observations show a single, relatively weak 37.7-GHz class~II methanol maser in this source with no accompanying emission in either the 38.3- or 38.5-GHz transitions.  The newly detected 36.2-GHz class~I methanol masers in this source are at the blueshifted end of the 6.7-GHz velocity range, in contrast to the 37.7-GHz emission which lies near the redshifted end.  Figure~\ref{fig:g348} shows that the class~II methanol masers are offset from the peak of the ATLASGAL dust clump by approximately 10 arcseconds, at the edge of a region of diffuse, weak infrared emission.  The strongest class~I methanol masers are closer to the peak of the dust continuum emission.  There is no radio continuum emission associated with any of the methanol maser sites, but there is a strong radio continuum source offset approximately 1 arcminute to the west of the masers (see Fig.~\ref{fig:g348}).  This radio continuum emission is associated an infrared source showing a strong excess in the 4.5-$\mu$m band.

\subsubsection{NGC6334F:} This is an iconic star formation region which has been widely studied, including the class~II methanol maser emission.  The 37.7-GHz methanol masers detected in the ATCA observations show a strong peak towards the blueshifted end of the spectrum, with weaker emission extending to -2.8~\kms\/, but no other distinct spectral peaks and no other emission stronger than 1~Jy (there is a small shoulder in the gradual declining strength of the emission at -8~\kms\/).  NGC6334F is unique in that the 38.3- and 38.5-GHz methanol masers are stronger than the 37.7-GHz masers.  This was also the case in the observations of \citet{Ellingsen+11a} and comparing the spectra of the three transitions between the two epochs (separated by approximately two years) we can see that the 37.7-GHz emission has decreased to approximately 70 percent observed by \citet{Ellingsen+11a}, while there is little difference in the 38.3- and 38.5-GHz emission.

Figure~\ref{fig:ngc6334f} shows the larger-scale environment around the NGC6334F star formation region.  The strongest class~II methanol masers are associated with an infrared source which shows strong radio continuum emission.  The 6.7-GHz methanol masers show strong emission from two sites, the second of which is offset by approximately 4 arcseconds north-east, in a region with no infrared emission.  The 37.7-, 38.3- and 38.5- GHz class~II methanol masers are all detected towards both of these sites and this is discussed in more detail in section~\ref{sec:classIIcomp}.  The 36.2-GHz class~I methanol masers in this source have previously been observed by \citet{Voronkov+14}, who find five class~I maser sites, four of which (A, B, C and E) have 36.2.GHz masers.  We detect 36.2-GHz emission towards three of the four sites - we do not find any convincing emission at site C, which is the weakest of the 36.2-GHz maser locations in \citet{Voronkov+14}.  We also detect strong 36.2-GHz methanol maser emission from a site outside the primary beam of the ATCA to the north (labelled F in Figure~\ref{fig:ngc6334f}.  This emission is associated with the well-studied class~I methanol maser source NGC6334I(N) \citep[e.g.][]{Kogan+98,Beuther+05b,Brogan+09}.

\subsubsection{G\,9.621$+$0.196:} This is the strongest 6.7-GHz methanol maser source and also shows periodic variability in a number of class~II methanol maser transitions \citep[e.g.][]{Goedhart+03}.  The ATCA observations confirm the speculation of \citet{Ellingsen+11a} that there is weaker 37.7-GHz maser emission covering the velocity of the strongest 6.7-GHz methanol masers (which peak at $+$1.3~\kms).  The current observations also detect very weak ($\sim$0.2~Jy) emission from the 38.3-GHz transition, but not from the 38.5-GHz transition.  Significant differences in the flux density of the 38.5-GHz masers compared to the 38.3-GHz masers has previously only seen towards G\,335.789$+$0.174, \citep{Ellingsen+13a}.

Figure~\ref{fig:g9} shows that the class~II methanol masers are located at the edge of an extended infrared source which also has associated radio continuum emission.  The class~II methanol maser site is close to several regions which show an infrared excess in the 4.5-$\mu$m wavelength range and the class~I methanol maser emission is predominantly to the east of the infrared, radio continuum and class~II methanol. 

\subsubsection{G\,23.440$-$0.182:} This is one of the weaker 37.7-GHz methanol masers and even at the greater sensitivity of the ATCA observations there is still only a single spectral component.  Although this may be a blend of several components at similar velocity, the low signal-to-noise does not allow us to be certain of this.  The relatively equatorial location of this source means that the synthesised beam of the ATCA is significantly elongated in declination.  Figure~\ref{fig:g23} shows the G\,23.440$-$0.182 star formation region - the methanol masers and radio continuum emission are located within an infrared dark cloud which has a number of embedded infrared sources.  The 37.7-GHz methanol masers are close to an infrared source which shows an excess in the 4.5-$\mu$m band.  In Fig.~\ref{fig:g23} we have used the synthesised beam as the restoring beam for the radio continuum observations, but in order to better localise the class~I methanol masers we have used a 3.2 arcsecond circular Gaussian beam as the restoring beam.  G\,23.440$-$0.182 has not previously been observed at 36.2-GHz and we can identify 5 sites.  The combination of the relatively strong class~I methanol masers and the elongated synthesised beam means there are artefacts apparent in the spectra shown in Fig.~\ref{fig:g23} due to interference between emission at different locations in velocity ranges that overlap.

\subsubsection{G\,35.201$-$1.736:} As discussed in Section~\ref{sec:res37} we do not detect any 37.7-GHz methanol masers at the epoch of these observations with a 5-$\sigma$ upper limit of 80~mJy.  We also do not detect any emission in either the 38.3- or 38.5-GHz transitions.  We do detect 36.2-GHz class~I methanol masers close to G\,35.201$-$1.736, although Figure~\ref{fig:g35} shows that they are significantly offset from the class~II methanol maser site and likely associated with a nearby, unrelated infrared source.  G\,35.201$-$1.736 lies outside the ATLASGAL and GLIMPSE survey regions and Fig.~\ref{fig:g35} shows a {\em WISE} three-colour infrared image as the background and there is strong infrared emission in the field which has saturated some of the bands.  G\,35.201$-$1.736 is located only a little more than a degree from declination zero and as an east-west oriented array, the synthesised beam of the ATCA observations is nearly one-dimensional for this source.  Because of this we have used a 4 arcsecond circular restoring beam for both the radio continuum and integrated 36.2-GHz methanol maser data presented in Fig.~\ref{fig:g35} and while the right ascension positions will be accurate, there is significant uncertainty in the declinations compared to the other sources observed.  Figure~\ref{fig:g35} suggests that the radio continuum peaks at the location of the infrared peak, with the class~II methanol masers, which here the position of the 6.7-GHz class~II methanol masers reported by \citet{Breen+15} are associated with an infrared point source with excess emission in the {\em WISE} 4.6-$\mu$m band which lies at the edge of the strong extended infrared source.  We identify three sites of class~I methanol masers west of the class~II masers and radio continuum emission, however, proper identification of the distribution of the class~I masers will require observations in an array configuration with north-south baselines.

\section{Discussion}

\subsection{Are the 37.7-, 38.3- and 38.5-GHz methanol masers co-spatial with the the 6.7-GHz?} \label{sec:classIIcomp}

The primary motivation for the current observations is to determine the degree to which the various class~II methanol maser transitions are co-spatial.  In the discussion below we use the terminology {\em maser spot} to refer to emission over a narrow velocity range from a compact region (typical velocity range of approximately 1~\kms\/ and angular scale less than 10 milliarcseconds).  The term {\em maser site} refers to a collection of one or more maser spots clustered on angular scales of a few hundred milliarcseconds and offset from other maser emission from the same transition by at least a few arcseconds. The first interferometric observations comparing the 6.7- and 12.2-GHz class~II methanol masers showed that some of the maser spots from these two transitions are co-spatial on milliarcsecond scales in some sources \citep{Menten+92,Norris+93,Ellingsen+96c}.  Analysis of the 12.2-GHz methanol masers associated with a statistically complete sample of 6.7-GHz methanol masers found that in 78 percent of sources where both transitions are observed the peak velocities are within 1~\kms\/ \citep{Breen+11}.  The demonstration that the two transitions are co-spatial in some sources, combined with the good agreement in the line of sight velocity of the peak emission in the majority, strongly suggests that the 6.7- and 12.2-GHz commonly arise from the same region (to within spatial scales of around 0.05 pc).  The degree to which this close association holds for the other class~II methanol maser transitions remains an open question, but one which the current observations can help to address.

\citet{Krishnan+13} used the ATCA to make observations of the 19.9-GHz ($2_{1} \rightarrow 3_{0}E$) transition towards 9 class~II methanol maser sources (and the 23.1-GHz $9_{2} \rightarrow 10_{1}\mbox{A}^+$ towards one source).  They found good agreement between the absolute positions of the 6.7- and 19.9-GHz class~II methanol maser sites ($\sim$0.2 arcseconds), but did not find any evidence for close correspondence between the maser spots from the two transitions.  One of the complicating factors in their investigation was that the 19.9-GHz methanol masers had relatively few spectral components (hence few spots) and in particular there were few sources with multiple strong components.  This makes comparison of the relative spatial distribution of the 19.9-GHz maser spots compared to the 6.7-GHz masers more difficult.  

The 37.7-GHz methanol maser transition is typically the fourth strongest of the class~II transitions (after the 6.7-, 12.2- and 107-GHz transitions) and hence potentially better for such comparisons.  Figure~\ref{fig:meth37} shows the spectra of the 37.7-GHz methanol masers and demonstrates that while the majority of the sources have multiple spectral components when observed with sufficient sensitivity, most have only a single strong component in the 37.7-GHz transition.  In addition, where there are multiple strong spectral components it is only possible to investigate the relative distribution when they have a spatial separation greater than around 0.1 arcseconds.  When measuring the position of the emission in each channel of the image cubes, we found that consecutive channels with emission stronger than approximately 1~Jy typically give good agreement (within a few 10s of milliarcseconds), whereas for weaker emission the scatter is larger (0.1 -- 1 arcsecond).  This is consistent with the expectation that the accuracy to which the relative position of emission in different spectral channels can be determined depends on the size of the synthesised beam and is inversely proportional to the signal-to-noise ratio.  So for the purposes of making comparison of the 37.7-GHz methanol masers with the 6.7-GHz distribution we focus on the components with a flux density greater than 1~Jy.

We have investigated the image cubes for all ten 37.7-GHz methanol maser detections and find that for six of the ten we do not have two or more 37.7-GHz maser components stronger than 1~Jy with a separation of more than 0.1 arcseconds (i.e. they fail one or both of the criteria identified above).  Hence for these six sources we cannot make a meaningful comparison between the 6.7- and 37.7-GHz emission.   The four sources for which a comparison can be made are G\,323.740$-$0.263, G\,339.884$-$1.259, G\,345.010$+$1.792 and NGC6334F (i.e. the four strongest 37.7-GHz methanol masers in the current sample).  For G\,323.740$-$0.263 we do not see any evidence for correspondence between the 6.7- and 37.7-GHz methanol masers (see Section~\ref{sec:indiv} for further details), whereas for the other three sources there is some agreement.

\begin{figure}
  \psfig{file=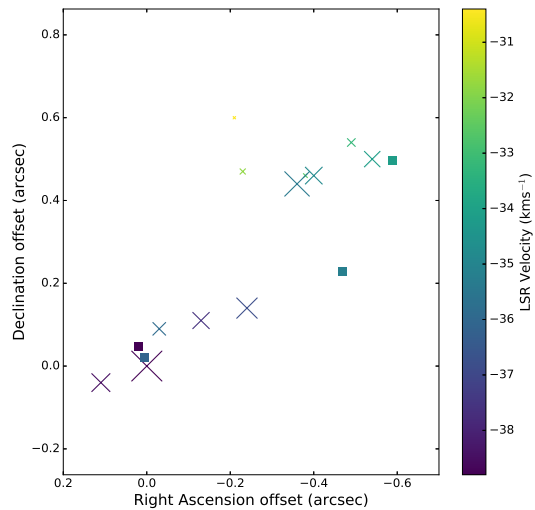,width=0.45\textwidth}
    \caption{The relative distribution of 6.7-GHz methanol masers determined by \citet{Walsh+98} (crosses) and 37.7-GHz methanol masers (squares) in G\,339.884$-$1.259.  For the 6.7-GHz methanol masers the size of the cross is proportional to the flux density of the component, while for the 37.7-GHz methanol masers all components are shown at the same size, as the strongest (-38.7~\kms) component is several orders of magnitude stronger than all the others.}
  \label{fig:g339_spot}
\end{figure} 

Figure~\ref{fig:g339_spot} shows a comparison of the 6.7- and 37.7-GHz methanol masers in G\,339.884$-$1.259.  The absolute positions have been shifted to better align the two transitions and we can see that the relative distribution is similar in terms of the orientation and velocity gradient of the emission.   The position angle of the elongation seen in the class~II methanol masers in this figure agrees with that seen on much larger scales in Figure~\ref{fig:g339} in the class~I methanol masers and dust continuum emission. 

\begin{figure}
  \psfig{file=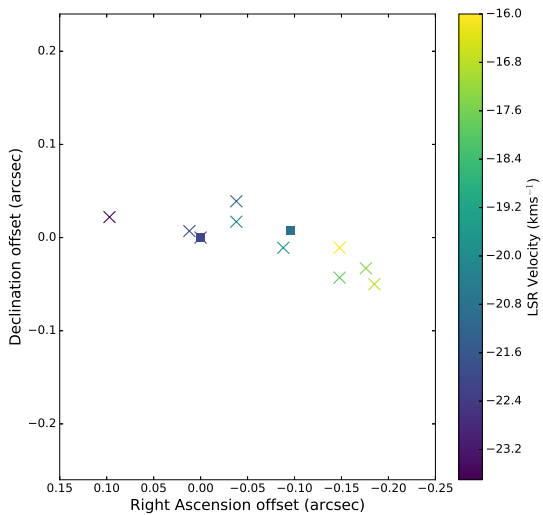,width=0.45\textwidth}
    \caption{The relative distribution of 6.7-GHz methanol masers determined by \citet{Norris+93} (crosses) and 37.7-GHz methanol masers (squares) in G\,345.010$+$1.792.}
  \label{fig:g345_spot}
\end{figure} 

Figure~\ref{fig:g345_spot} shows a comparison of the 6.7- and 37.7-GHz methanol masers in G\,345.010$+$1.792.  Here we only detect two components at 37.7-GHz and the absolute positions have been shifted to align the -22.4-\kms\/ component from the \citet{Norris+93} 6.7-GHz distribution with the -22.0-\kms\/ component at 37.7-GHz.  We can see that when we do that the secondary component at 37.7-GHz is offset in the same general direction and sense of velocity gradient as observed at 6.7-GHz, but that  there is a difference in the velocity of the 37.7-GHz secondary component and the nearest 6.7-GHz counterpart.

\begin{figure}
  \psfig{file=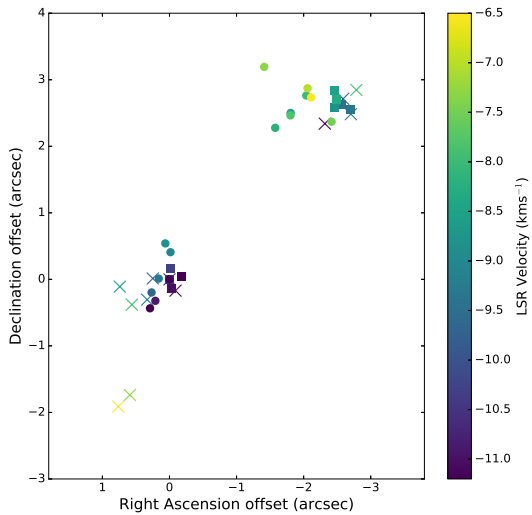,width=0.45\textwidth}
    \caption{The relative distribution of 6.7-GHz methanol masers determined by \citet{Norris+93} (crosses) with the 37.7-GHz methanol masers (squares) and 38.3-GHz methanol masers (circles) in NGC6334F.  The 37.7- and 38.3-GHz positions have been determined from ATCA data including all antennas, which gives a synthesised beam of approximately 2.3 $\times$ 1.4 arcseconds.}
  \label{fig:ngc6334f_spot}
\end{figure} 

Figure~\ref{fig:ngc6334f_spot} shows a comparison of the 6.7- and 37.7-GHz methanol masers in NGC6334F.  NGC6334F differs from all other sources in our sample in that the 37.7-GHz methanol maser emission clearly arises from more than one site.  The observations of \citet{Norris+93} and \citet{Ellingsen+96c} show 6.7-GHz methanol maser emission from 3 separate sites with offsets of a few arcseconds between each.  We reliably detect 37.7-GHz emission over a wide velocity range, however, because the separation of the different class~II methanol maser sites is comparable to the size of the synthesised beam of our observations (when using data from only the inner 5 antennas), where two or more sites show emission at the same velocity, we measure the position as the flux-density weighted average of the two sites.  The result is that, rather than showing emission at two (or three discrete sites), these data show emission distributed along a line between the two main 6.7-GHz methanol maser sites, and also emission in a similar elongated structure to the north-east of the northern-most 6.7-GHz site.  To minimise this issue, we imaged the 37.7-, 38.3- and 38.5-GHz methanol masers in NGC6334F using the data from all ATCA antennas which yields a synthesised beam of approximately 2.3 $\times$ 1.4 arcseconds.  These data have a large gap in the {\em uv-}coverage, however, because the morphology of the maser emission is point-like we are still able to obtain reasonable images.  At this higher angular resolution the 37.7-, 38.3- and 38.5-GHz methanol masers can clearly be seen to originate in two clusters, associated with the same regions which show the strongest 6.7- and 12.2-GHz methanol masers in  Figure~\ref{fig:ngc6334f_spot}.  

\begin{figure*}
  \psfig{file=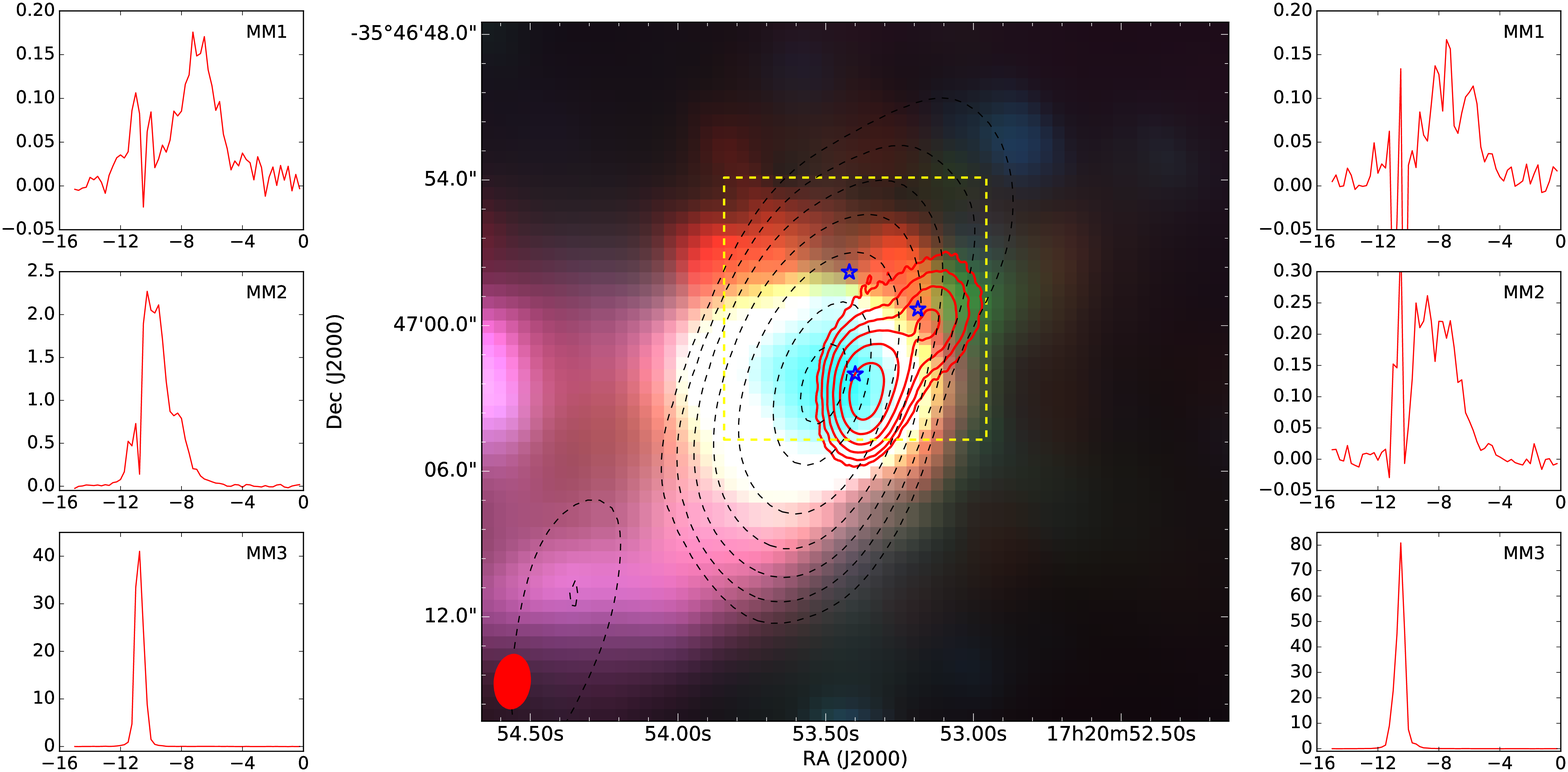,width=0.90\textwidth}
  \psfig{file=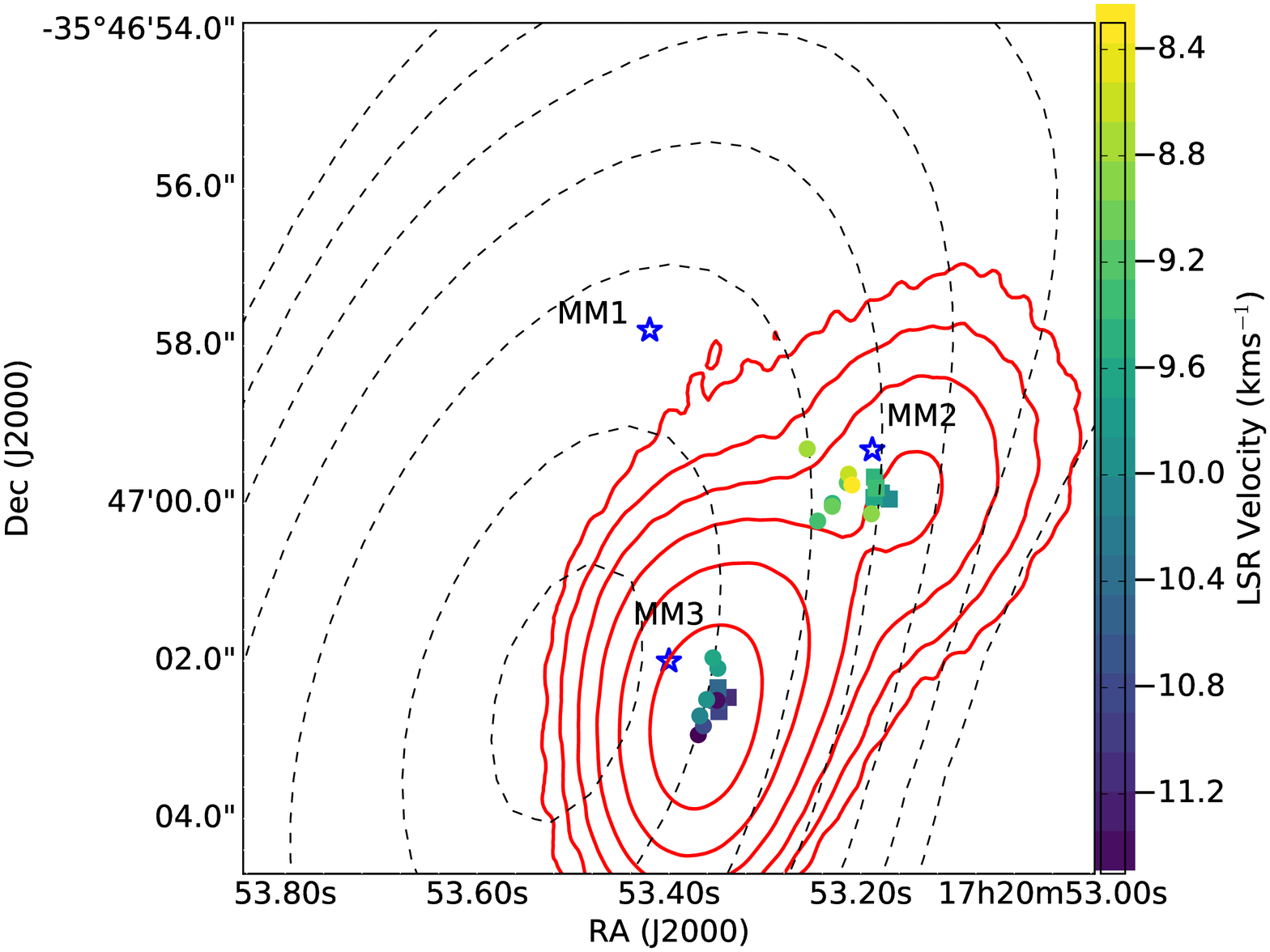,width=0.58\textwidth}
    \caption{{\em Top Panel:} The background image is from {\em Spitzer} IRAC observations from the GLIMPSE survey \citep{Benjamin+03} of the NGC6334F region with blue, green, and red from the 3.6, 4.5, and 8.0 $\mu$m bands, respectively. The black dashed contours are the 8-mm radio continuum emission from the current ATCA observations (inner 5 antennas), with contours at 2, 4, 8, 16, 32 and 64 and 90 percent of 2.01 Jy beam$^{-1}$.  The red contours are the integrated 37.7-GHz methanol maser emission utilising data from all ATCA antennas with contours at 2, 4, 8, 16, 32 and 64 percent of 29.4 Jy \kms\/ beam$^{-1}$.  The blue stars mark the locations of the millimetre sources MM1, MM2 and MM3, (from north to south, respectively), identified by \citet{Brogan+16}.   The spectra to the left of the image show the 37.7-GHz methanol maser emission extracted from the image cube at the specified locations and the spectra to the right show the 38.3-GHz methanol maser emission from the same locations.  The red-filled ellipse in the bottom-left corner represents the synthesised beam of the 37.7-GHz observations. The x-axis of the spectra are line-of-sight velocity in \kms\/ and the y-axis of the spectra show the flux density in Jy.  {\em Bottom Panel:} The region of the top image marked by the yellow-dashed square.  The black-dashed and red contours and the blue stars are the same as in the top image, the squares and circles show the distribution of the 37.7- and 38.3-GHz methanol masers, respectively, from Fig.~\ref{fig:ngc6334f_spot}.}
  \label{fig:ngc6334f_classII}
\end{figure*}

Figure~\ref{fig:ngc6334f_classII} shows the location and extension of the integrated 37.7-GHz methanol maser emission with respect to the radio continuum, infrared emission and location of millimetre continuum sources.  The two main sites of 6.7-GHz methanol masers are towards the continuum sources labelled as MM3 (the well known ultra-compact \ionhy region which which has associated radio continuum emission) and MM2 (which does not show centimetre radio continuum) \citep{Brogan+16}.  Figure~\ref{fig:ngc6334f_spot} shows that the 38.3-GHz maser distribution associated with the MM3 region extends north-west from that observed at 6.7-GHz and for the emission from the MM2 region it is displaced to the east.  The 37.7-GHz masers show the same trend, but to a lesser extent.  This is almost certainly due to small degree of residual spectral/spatial blending between different sites for these transitions.  Unfortunately, for the 37.7-, 38.3- and 38.5-GHz transitions there is only a single strong maser component, with all the emission towards the MM2 site being much weaker (around 1~Jy at 37.7-GHz and $<$ 0.5~Jy in the other two transitions).  This means that it is not possible with the current data to rigorously test whether the relative distributions of any of these class~II maser transitions show close correspondence with the 6.7- and 12.2-GHz masers.  We also imaged the 38.5-GHz methanol masers at higher angular resolution, the results are very similar to those observed for the 38.3-GHz transition.  To avoid clutter and confusion within the images we show only the 38.3-GHz results, but the comments below relating to the spatial distribution of the 38.3-GHz methanol masers in NGC6334F apply equally to the 38.5-GHz transition.

The offset of the 38.3- and 37.7-GHz emission north-east of the 6.7-GHz masers in MM2 suggests the presence of additional 37.7- and 38.3-GHz emission not associated with a known 6.7-GHz methanol maser site.  Figure~\ref{fig:ngc6334f_classII} shows the distribution of the 37.7-GHz methanol emission with respect to the \citet{Brogan+16} millimetre sources MM1, MM2 and MM3.  The offset of the MM2 38.3- and 37.7-GHz methanol emission is in the general direction of MM1 and the integrated emission shows some extension in this general direction.  The lower-resolution ATCA data is more sensitive to larger scale emission and shows stronger relative emission towards the MM1 site.  The spectra in Fig.~\ref{fig:ngc6334f_classII} are extracted from the image cubes of the self-calibrated 37.7-GHz (left-hand side) and 38.3-GHz (right-hand side) methanol data.  The strongest emission in both transitions is towards the MM3 site and because the angular offset of the MM2 and MM1 sites from MM3 is comparable to the synthesised beam of the current observations, artefacts due to the strong peak are observed at that velocity at the offset locations.  Toward MM2 we can see that at 37.7-GHz there is additional emission with flux density of approximately 2~Jy, present at redshifted velocities (around -8~\kms) and weaker emission in the same velocity range at 38.3-GHz.  While towards MM1 there seems to be a broad, weak spectral component present in both transitions.  This strongly supports the hypothesis that the 37.7-, 38.3- and 38.5-GHz methanol emission is located at three sites within NGC6334F, associated with MM3, MM2 and MM1.  We would expect that observations with higher angular resolution (0.5 arcsecond or better) would localise the 37.7- and 38.3-GHz emission towards MM3 and MM2 and show better agreement with the 6.7- and 12.2-GHz results.  The emission associated with MM1 may be thermal, or quasi-thermal emission, rather than maser emission and so may be resolved out in higher resolution observations.  The 19.9- and 23.1-GHz class~II transitions in NGC6334F have also been imaged using the ATCA, however, for those transitions emission is only observed towards the MM3 site \citep{Krishnan+13}.

In summary, from our imaging data we observe good agreement in the absolute positions (i.e. within the expected astrometric accuracy), and a broad similarity in the distribution (for three of the four sources where this can be investigated).  This is good evidence that in general the 37.7-GHz methanol masers arise from the same sites as the 6.7- and 12.2-GHz masers (and likely all the other class~II transitions).  However, with the current data we are not able to test whether some of the individual 37.7-GHz maser spots are co-spatial with those at 6.7-GHz and 12.2-GHz.  For most sources this is because there are not enough strong spectral components in the 37.7-GHz emission to enable rigorous comparison with other class~II maser distributions.  The best prospect for a more definitive test appears to be higher angular resolution observations of the 37.7-GHz methanol masers in the NGC6334F star formation region, as emission is observed over a wide velocity range at multiple sites.

One additional test that can be made with the current data is to make a detailed comparison of the spectra of the 6.7- and 37.7-GHz masers.  To demonstrate that emission is co-spatial between two transitions it is necessary to show both that the emission arises from the same location at high angular (spatial) resolution, and also to demonstrate agreement in the spectral characteristics (peak velocity, line shape etc).  The individual maser spots arise when there is a sufficiently long path through the region of masing gas, that has a high degree of line-of-sight velocity coherence in the specific direction of the observer.  This does not require the entire path to share a specific line-of-sight velocity to the observer, just a sufficient fraction of that particular path.  The physical conditions within the region where the masers arise will not be completely uniform, so since the various maser transitions have different sensitivities with respect to changes in the physical conditions we expect different parts of the velocity coherent line-of-sight to the observer will favour different transitions.  The net result of this is that two transitions can have co-spatial maser spots in the plane of the sky, but still exhibit differences in the line profile due to differences in the line-of-sight contributions.  This will always be the case to some degree, however, similarities (and differences) in the spectral profiles provide some independent information as to the degree to which the emission might be co-spatial between transitions.  

That the 6.7- and 12.2-GHz methanol masers have the same peak velocity to within 1~\kms\/ in 78 percent of sources (within 0.2~\kms in 56 percent) is consistent with a high degree of co-spatial emission between the two transitions \citep{Breen+11}.   Comparing the peak velocity of the 37.7- and 6.7-GHz methanol masers we find that they have the same peak velocity to within 1~\kms\/ in 7 of 10 sources, comparable to the 6.7- and 12.2-GHz result.  Only 3 of 10 sources have the peak velocities agreeing to within 0.2~\kms, but for such a small sample this is not significantly different to the 6.7- and 12.2-GHz results.  Figure~\ref{fig:compare} shows a comparison between the 37.7-GHz and 6.7-GHz methanol maser emission in the ten sources detected in the current observations.  The 6.7-GHz methanol maser data are the so-called MX spectra from the methanol multibeam survey (MMB), which were observed with the Parkes telescope in 2008 and 2009 \citep{Green+10,Caswell+10,Caswell+11,Green+12a,Breen+15}.  To facilitate the comparison between the spectral components the 6.7-GHz methanol maser spectra have been scaled so that it has the same peak flux density as the 37.7-GHz methanol emission.  The 37.7-GHz methanol maser data is that extracted from the image cubes, which have a 0.25~\kms\/ velocity resolution (note that the intrinsic resolution of the observations is slightly coarser than this at 0.3~\kms).  The 6.7-GHz methanol maser spectra have a spectral resolution of 0.11~\kms.

Figure~\ref{fig:compare} shows that for the majority of the sources there is not particularly good alignment of the 37.7-GHz peaks with a 6.7-GHz methanol maser spectral feature.  Often the strongest emission is at similar velocities for the two transitions, however, in most cases the peak is offset by a greater amount than can be accounted for by the difference in the velocity resolution.  Furthermore, the line profiles are in many cases sufficiently different (e.g. G\,323.740$-$0.263, G\,337.705$-$0.053) that it is clear that velocity resolution is not solely responsible for the differences in the peak velocity.  The two sources where we see the best alignment of the 37.7-GHz methanol maser peak with a 6.7-GHz counterpart are the sources G\,339.884$-$1.259 and G\,348.703$-$1.043.  G\,339.884$-$1.259 is one of the sources where we find broad similarities in the spatial distribution of the 37.7-GHz methanol masers compared to the 6.7-GHz and interestingly the other two sources where  a broad similarity in the spatial distributions is observed  (G\,345.010$+$1.792 and NGC6334F) also show better alignment than the majority (particularly G\,345.010$+$1.792).  

General class~II methanol maser modelling \citep{Sobolev+97b,Cragg+05} shows that the 37.7-GHz transition is quite sensitive to the physical conditions of the associated gas.  Examination of table~1 of \citet{Sobolev+97b} shows that the 37.7-GHz transition appears to favour cool (30 K), dense (10$^{7}$ cm$^{-3}$) gas with warm dust and the presence of background continuum radiation.  Furthermore, table~2 of \citet{Cragg+05} shows that many of the rarer class~II methanol maser transitions are only present in cool gas.  These investigations also show that in general it is difficult to have a strong 37.7 GHz maser without even stronger 6.7 GHz from the same volume of gas.  However, it is important to note that the conditions which produce the brightest 6.7- and 12.2-GHz methanol masers do not correspond to the conditions that produce the brightest 37.7-GHz methanol masers.  So maser pumping models suggest that while the 37.7-GHz methanol masers should arise from regions which show stronger 6.7- and 12.2-GHz methanol masers, we do not necessarily expect the strongest 37.7-GHz methanol masers to correspond to the strongest 6.7- and 12.2-GHz maser components.  We have examined the intensity of the 6.7- and 12.2-GHz methanol maser spectra at the velocity of the 37.7-GHz methanol maser peak and find that for all but one source (G\,345.010$+$1.792), the 6.7 GHz methanol maser emission has a flux density a factor of three or  more higher than the 37.7-GHz flux density at the velocity of the 37.7-GHz peak emission.  For the 12.2-GHz emission there are two sources for which the 37.7-GHz peak is stronger than the 12.2-GHz maser emission at the same velocity (G\,337.705$-$0.053 \& G\,9.621$+$0.196).  

The 6.7-GHz methanol maser spectra for the MMB were primarily observed in 2008 and 2009, several years prior to the 37.7-GHz observations, as were the 12.2-GHz methanol maser follow-up observations of \citeauthor{Breen+12a}.  So variability can potentially affect the comparison of the spectra of different transitions toward the same source, however, the MMB observations and comparison with earlier data shows that peak velocity of strong 6.7-GHz methanol masers generally does not change greatly on timescales of a few years.  So it is unlikely that variability is playing a significant role in the differences observed between the transitions.  Combining our investigation of the spatial distribution of the 37.7-GHz methanol masers and the spectral comparison we conclude that while the 37.7-GHz methanol masers may arise from the same sites as the 6.7-GHz methanol masers they are less often, or to a lesser degree co-spatial than the 12.2-GHz masers are with the 6.7-GHz transition.  We note that in agreement with theoretical predictions, our observations suggest that the brightness temperatures of the 37.7-GHz masers are always lower than those of the 6.7-GHz and 12.2-GHz methanol masers at the same velocity.  However, sensitive observations made at the same epoch with higher spectral resolution (particularly for the 37.7-GHz spectra) are required to provide definitive data for testing the degree of co-spatiality and relative brightness temperatures.

\begin{figure*}
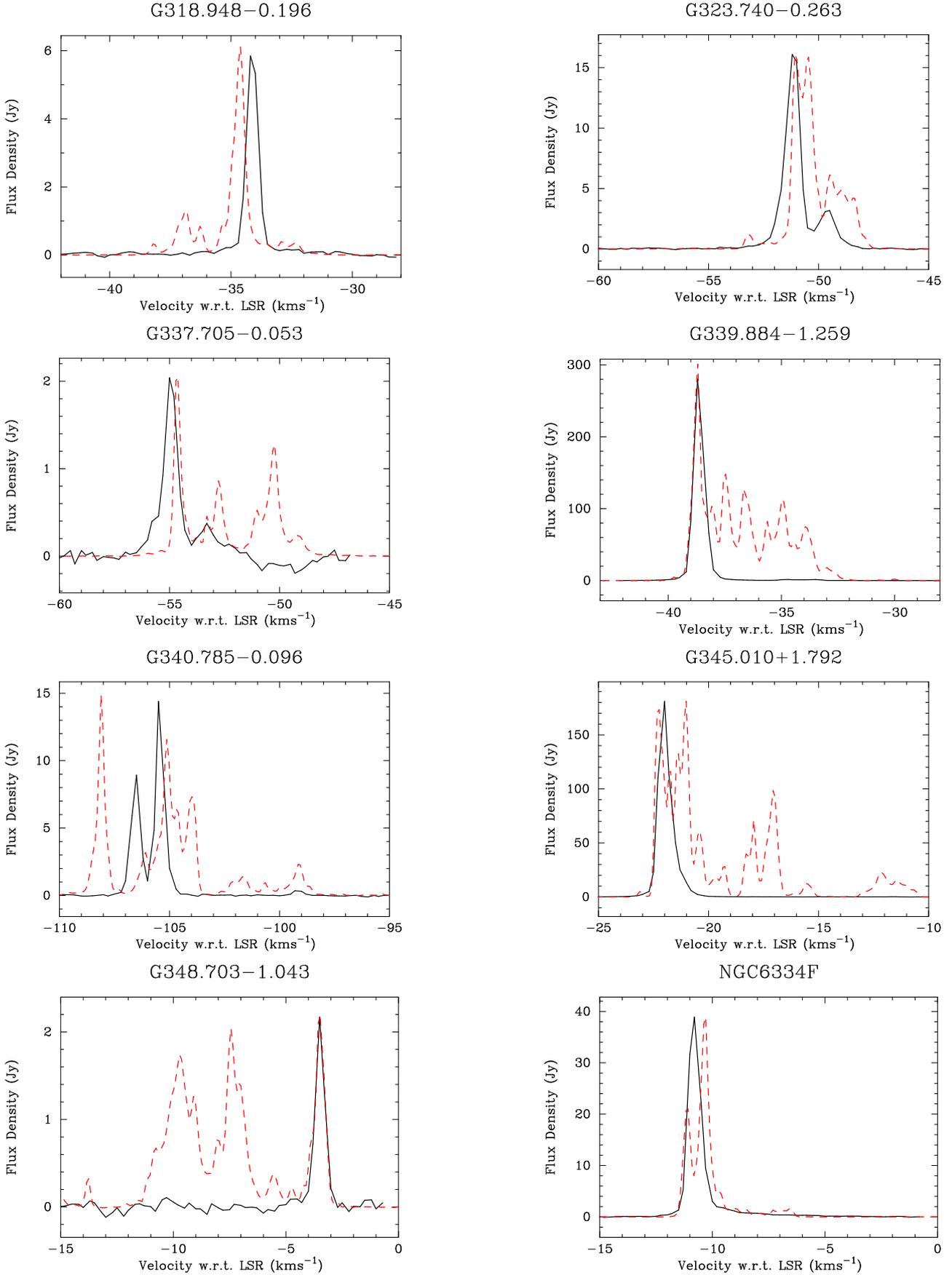

  \specdfig{g318.948_comp}{g323.740_comp}
  \specdfig{g337.705_comp}{g339.884_comp}
  \specdfig{g340.785_comp}{g345.010_comp}
  \specdfig{g348.703_comp}{ngc6334f_comp}
  \caption{Spectra of the 37.7-GHz methanol maser emission (solid line), compared to the 6.7-GHz methanol masers scaled to have the same peak flux density as the 37.7-GHz emission (red dashed line).}
  \label{fig:compare}
\end{figure*} 

\begin{figure*}
  \specdfig{g9.621_comp}{g23.440_comp}
  \contcaption{}
\end{figure*} 

\subsection{The peak velocity of 37.7-GHz methanol masers} \label{sec:vel}
\begin{figure*}
  \psfig{file=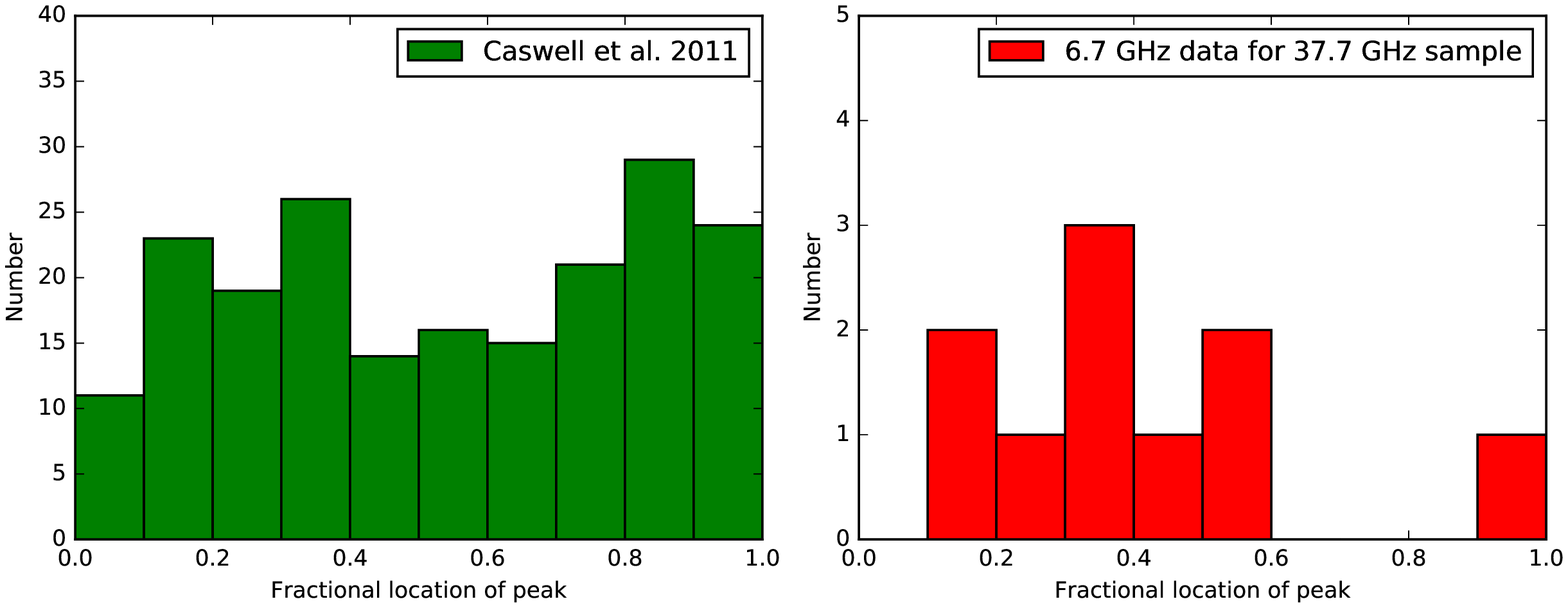,width=0.90\textwidth}
    \caption{The location of the peak of the 6.7-GHz methanol maser emission as a fraction of the total velocity range of the 6.7-GHz emission.  The left-hand plot shows the distribution for the 6.7-GHz methanol masers in the \citet{Caswell+11} paper (198 sources).  The right-hand plot shows the same information (location of the 6.7-GHz methanol maser peak within the 6.7-GHz velocity range) for the 37.7-GHz masers.}
  \label{fig:velhist}
\end{figure*}

Figure~\ref{fig:compare} shows that the 37.7-GHz methanol maser peaks are preferentially located towards the blueshifted (more negative velocity) end of the 6.7-GHz methanol maser range.  From the ten sources in our sample the only exceptions are G\,$318.948-0.196$, G\,$348.703-1.043$ and to a lesser extent G\,$340.785-0.096$.  In total 7 of the 10 sources have the 37.7-GHz methanol maser peak velocity in the blueshifted half of the total 6.7-GHz velocity range.  It has previously been noted that some of the most luminous 6.7-GHz methanol masers have peak velocities towards the blueshifted end of the velocity range (e.g. G\,339.884$-$1.259, G\,345.010$+$1.792 and NGC6334F), however, there are also examples of sources where this is not the case (e.g. G\,318.948$-$0.196 and G\,9.621$+$0.196).  \citet{Sobolev+05} and \cite{Kirsanova+17} have investigated the difference between the mid-velocity of the 6.7-GHz masers and the systemic velocity of the thermal gas (as traced by CS ($J=2-1$) emission), in two different longitude regions.  They find that there is a preference for the methanol masers to be blueshifted with respect to the thermal gas in the Galactic Longitude range $l$ = 320\degrees -- 350\degrees, but redshifted in the range $l$ = 84\degrees -- 124\degrees.  They interpret this in terms of large-scale biases in the location of the masers relative to the spiral arms.  The majority of the sources we are considering here are in the longitude range where \citet{Sobolev+05} find a blueshifted bias, so it is possible that large-scale Galactic structure is playing some role here.  To provide a robust baseline for comparison we have  investigated the relative location of the velocity of the 6.7-GHz peak within the total velocity range for the 198 sources in the \citet{Caswell+11} MMB paper (Galactic longitude range $l$ = 330\degrees -- 345\degrees) and find that it is approximately uniform (Figure~\ref{fig:velhist}, left-hand panel).  For this sample of sources 93 (47 percent) are within the blueshifted half of the velocity range.  If we collate the same information from the MMB survey for the 6.7-GHz masers associated with the ten 37.7-GHz detections from the current observations, we find that 70 percent are in the blueshifted half of the velocity range (Figure~\ref{fig:velhist}, right-hand panel).  This suggests that the blueshifted tendency for 37.7-GHz methanol masers is significantly stronger than that for the larger population of class~II methanol masers.

If we consider the location of the 37.7-GHz methanol maser peak velocity within the 37.7-GHz velocity range the results are more extreme, with 8 of the 10 sources within the blueshifted half of the velocity range.  Furthermore, 0 of the 10 sources are in the most redshifted 30 percent of the range and 6 of the 10 sources in the most blueshifted 30 percent of the range.  If we assume that the location of the peak velocity of the 37.7-GHz methanol masers within the velocity range is uniform, then the probability of 6 or more from a sample of 10 sources being within the most blueshifted 30 percent is less than 5 percent, while the chance of 0 from 10 being within the redshifted range is less than 3 percent.  So it is unlikely that the peak of the 37.7-GHz methanol masers has an equal chance of being anywhere within the velocity range of the emission, our data provide good evidence that the peak is preferentially towards the blueshifted half of the range.  Furthermore, the class~II methanol masers with an associated 37.7-GHz methanol maser are more likely to have a peak in the blueshifted half of the velocity range than the overall population.

How does the class~II methanol maser peak and velocity range compare or relate to the general molecular gas in a region?  A number of studies have shown that both the peak and mid-range velocity of class~I methanol maser sources are an excellent proxy for the systemic velocity of the region \citep[e.g.][]{Voronkov+14,Jordan+17,Yang+17}.  We have made a comparison between the peak velocity of the 37.7-GHz class~II methanol masers and the 36.2-GHz class~I methanol masers and find that for eight of the ten sources from the current sample, the class~II methanol maser peak is blueshifted compared to the class~I methanol maser peak.  We performed a similar analysis comparing the velocity of the 36.2-GHz methanol maser peak for each source observed by \citet{Voronkov+14}, with the velocity of the 6.7-GHz methanol maser peak measured in the MMB survey.  \citet{Voronkov+14} observed 71 36.2-GHz methanol maser sources and 62 of these have a 6.7-GHz methanol maser within 1 arcminute (where there was more than one 6.7 GHz methanol masers with 1 arcminute, the source with the smallest angular separation was assumed to be associated).  For this sample of 62 sources 39 (62 percent) have the class~II methanol maser emission redshifted with respect to the class~I methanol maser peak velocity.  So once again, the 37.7-GHz methanol maser sources show very different behaviour to larger, more representative samples.  The tendency for the 37.7-GHz methanol maser peak velocity in our observations  to be blueshifted compared to the systemic velocity suggests that they are preferentially observed from molecular gas outflowing with respect to the systematic velocity and directed towards the observer.

\subsection{Brightness temperature of the class~II masers} \label{sec:brightness}
The current ATCA observations represent the highest angular resolution observations undertaken of the 37.7- 38.3- and 38.5-GHz class~II methanol maser transitions.  Although we have only used the data for the inner-five antennas for the majority of the analysis, the data to the sixth antenna was recorded.   Analysis of the longer baseline data for the strongest component in all the masers with a peak flux density greater than 5~Jy (i.e. 7 of the 10 sources at 37.7-GHz and 2 sources at 38.3-/38.5-GHz) shows that in all cases the emission is unresolved on the longest baselines (approximately 4.4~km).  This corresponds to a {\em uv}-distance of approximately 540~k$\lambda$, or an angular scale of $\sim$0.38 arcseconds.  

The strongest 37.7-GHz methanol maser (G\,339.884$-$1.259) has a peak flux density of $\sim$300~Jy and so we can set a lower limit on the brightness temperature of that maser component of $2.6 \times 10^6$~K and a lower limit on the brightness temperature of G\,$318.948-0.196$ (peak flux density of 6.1~Jy) of $5.3 \times 10^4$~K.  The strongest 38.3- and 38.5-GHz methanol maser emission is detected towards NGC6334F, which has a peak flux density of $\sim$120~Jy and $\sim$150~Jy for the two transitions, respectively.  This corresponds to a lower limit on the brightness temperature for these two transitions of $9.9 \times 10^5$~K and $1.2 \times 10^6$~K respectively.  Brightness temperatures in excess of 10$^{11}$~K are observed for some 6.7- and 12.2-GHz methanol masers \citep[e.g.][]{Menten+88a,Menten+92}.  

The class~II methanol masers in NGC6334F have been modelled by \citet{Cragg+01} who predict brightness temperatures for the 38.3- and 38.5-GHz class~II methanol masers between $\sim 10^{-3}$ -- $10^{-7}$ times that of the 6.7-GHz masers ($>$ 10$^{10.1}$~K).  While for the 37.7-GHz methanol masers the prediction is a brightness temperature of a factor of $\sim 10^{-4}$ -- $10^{-6}$ lower.  The current observations are broadly consistent with the modelling of \citet{Cragg+01}, with the ratio of the lower limit of the 37.7-, 38.3- and 38.5-GHz methanol maser brightness temperatures compared to the 6.7-GHz lower limit being $10^{-4.6}$, $10^{-4.1}$ and $10^{-4.0}$, respectively.  So the observations appear likely more consistent with brightness temperature values  at the higher end of the modelled range.  \citet{Cragg+01} also modelled the class~II methanol maser emission for G\,345.010$+$1.792, however, at the time of that study the 37.7-, 38.3- and 38.5-GHz masers had not been observed in this source and so the predictions for those transitions are not given.  The current observations show that the ratio of the lower limit of the 6.7-GHz masers ($>$ 10$^{11.5}$~K) to the lower limit of the 37.7-, 38.3- and 38.5- GHz methanol masers to be $10^{-5.3}$, $10^{-6.6}$ and $10^{-6.8}$ respectively  for G\,345.010$+$1.792, quite different from what is observed for NGC6334F.

\subsection{Evolutionary phase of 37.7-GHz methanol masers}
A focus of a number of different large-scale maser surveys has been to use the results to established a statistically-based evolutionary timeline for masers associated with high-mass star formation regions \citep[e.g.][]{Breen+11,Voronkov+14,Titmarsh+16,Jordan+17}.  These investigations are aiming to quantify the qualitative evolutionary timelines that have been proposed \citep{Ellingsen+07,Breen+10a}.  Where rare, weak maser transitions such as the 37.7-GHz methanol masers fit within such schemes has been discussed in some detail previously and we will not repeat that discussion here \citep{Ellingsen+11a,Ellingsen+13a,Krishnan+13}.  \citeauthor{Ellingsen+11a} showed that 37.7-GHz methanol masers are only found in star formation regions with the highest 6.7- and 12.2-GHz methanol maser luminosities and suggested that that indicated they arise towards the end of the class~II methanol maser phase.  

Here we investigate the new information provided by the more sensitive high-resolution observations and its implications for the evolutionary timeline.  Some of the analysis undertaken requires the distance to the source to be known or estimated and we have tabulated this information in last column of Table~\ref{tab:res36}.  For 5 of the 11 target sources we have trigonometric parallax measurements of the distance to the sources, for the remaining 6 sources we have used the Baysian distance calculator of \citet{Reid+16}, assuming the near kinematic distance when that is the primary means by which the distances have been inferred.  For some of the sources the kinematic distance estimates are a little different from those used in previous works, as we have used the peak velocity of the 36.2-GHz methanol masers as the line-of-sight velocity, as that is a more reliable estimate of the systemic velocity (see discussion in Section~\ref{sec:vel}), however, it differs from the class~II methanol maser peak (and mid-point) velocity.

The 37.7-GHz methanol maser transition is rare and if, as hypothesised, it traces a short-lived evolutionary phase, the presence of two distinct sites of such masers in NGC6334F is not expected.  The strongest class~II methanol masers in the NGC6334F region are associated with the well-known cometary \ionhy region (and millimetre source MM3), with a second site of strong class~II methanol maser emission offset approximately 4 arcseconds to the north-west.  This second site does not have an associated infrared source and has only been detected at millimetre wavelengths \citep[source MM2 ;][]{Hunter+06,Brogan+16}.  However, both class~II methanol maser sites show strong 6.7- and 12.2-GHz methanol masers (6.7-GHz peak flux densities of around 1000~Jy), so each independently meet the criteria identified by \citet{Ellingsen+11a} as being likely to host a 37.7-GHz methanol maser.  The lack of an infrared and radio counterpart to MM2 means that it is a highly embedded source, suggesting it is younger than MM3, which has a strong ultracompact \ionhy region and OH masers \citep{Forster+99,Brooks+01}.   The two class~II methanol maser sites have similar millimetre flux densities, but \citet{Hunter+06} estimate that MM3 is the more massive young stellar object. Overall, \citet{Hunter+06} find 4 high-mass young stellar objects in the NGC6334F region and one possibility for the presence of multiple sites of 37.7-GHz methanol maser emission is that in dense clusters it can arise due to evolution of the gas and dust in the system, rather than that associated with a single young stellar object. 

\citet{Voronkov+14} found that the linear spread of class~I methanol maser features and the total velocity width of the class~I emission was significantly larger for those which had an associated OH maser compared to those which did not (see figure~5 of their paper).  In their sample, the class~I methanol masers without an associated OH maser all had spatial extents $<$ 0.4~pc and velocity ranges $<$ 10~\kms, whereas the majority of OH-associated sources exceeded one or both of these limits.  All of the eleven target sources for the current observations have an associated ground-state OH maser \citep[see table~4 of][]{Ellingsen+13a}, however, we exclude G\,$35.201-1.736$ from consideration, as the class~I methanol masers don't appear to be associated with the class~II masers (see comments for this source in Section~\ref{sec:indiv}).  We measured the linear spread of the class~I masers by finding the angular separation between the most widely separated 36.2-GHz components for each source, using the positions recorded in Table~\ref{tab:components} and converting to a linear scale using the distance estimates from Table~\ref{tab:res36}.  We find four of the sources (G\,318.948$-$0.196, G\,339.884$-$1.259, G\,340.785$-$0.096 and G\,23.440$-$0.182) have a linear spread $>$0.4~pc, with a further two sources having velocity widths greater than 10~\kms\/ (G\,$337.705-0.053$ and G\,$9.621+0.196$).  So in total 6 of 10 sources have either a linear spread $>$0.4~pc (4 sources), a velocity width $>$10~\kms\/ (4 sources), or both (2 sources).  It should be noted that 5 of the 11 sources here are included in the \citet{Voronkov+14} sample (and are all OH-associated sources) and that observations of \citeauthor{Voronkov+14} included both the 36.2- and 44-GHz methanol maser transitions and the latter of these is generally more common and stronger in high-mass star formation regions.  This means that the linear extents measured in the current observations are likely systematically lower than those in \citet{Voronkov+14} and hence the current observations are consistent with the class~I methanol masers being associated with more evolved star formation regions.

One of the outstanding questions for the evolutionary timeline for masers is where class~I methanol masers fit into the scheme.  The well-established association of many class~I methanol maser sources with outflows \citep{Voronkov+06,Chen+09,Cyganowski+09}, would seem to favour an association with younger star formation regions, where outflows are expected to be stronger and more prevalent.  However, there are numerous examples of class~I methanol masers associated with star formation regions where OH masers are observed and these have been proposed to be older regions in the evolutionary timeline \citep{Breen+10a}.  The presence of class~I methanol masers towards all the sources in the current 37.7-GHz methanol maser sample - a transition which is hypothesised to arise just prior to the end of the class~II methanol maser stage, is further evidence that some class~I methanol maser are associated with older more evolved star formation regions.  The figures in Appendix~\ref{sec:environ} show that where there is continuum emission associated with the class~II methanol maser site, the class~I masers are typically distributed around it, again suggesting an association with a more evolved star formation region.  However, it should be noted that sensitive observations, such as those reported here, detect class~I methanol masers towards a very high percentage of high-mass star formation regions.  The MALT-45 survey of \citet{Jordan+17} made a blind survey of a 5 square degree region of the Galactic Plane which detected 77 class~I methanol maser sources (in the 44-GHz transition).  Of these 77 masers, 31 are associated with a 6.7-GHz class~II methanol maser, showing that more than half of the class~I maser sources are not associated with a class~II maser.  \citet{Jordan+17} find that the majority of sources have a linear extent $<$0.5~pc and a velocity width $<$5~\kms\/ and also that those class~I methanol masers which are not associated with OH masers or radio recombination lines are generally less luminous than those which are.  The class~I methanol masers are typically associated with dust clumps with masses $>$ 1000\msol, suggesting the majority are associated with high-mass star formation regions.  From this they infer that the majority of the class~I methanol masers without an associated class~II methanol maser are younger sources.  

At present there are approximately 500 known Galactic class~I methanol maser sources \citep[see ][and references therein]{Valtts+10,Yang+17}, less than half the number of known class~II methanol masers.  However, this number will significantly increase in the near future with the expansion of the MALT-45 survey area to 90 square degrees as part of an ATCA Legacy science project.  It is likely that the total number of known class~I methanol maser sources in the Galaxy (distinct regions showing class~I maser emission), will soon exceed the class~II methanol maser population.  This suggests that the either (or both) of the duration, or mass range for the class~I methanol maser phase of the evolutionary timeline exceeds that of the class~II methanol masers.  Class~I methanol masers have been detected towards lower-mass stars \citep[e.g.][]{Kalenskii+10}, however, the association of the vast majority of the  new \citet{Jordan+17} class~I maser detections with high-mass dust clumps suggests that the number associated with high-mass star formation regions alone will exceed the class~II population. So if the primary reason for more class~I methanol masers is the duration of this phase, then this is consistent with them being present in regions prior to the onset of class~II methanol maser emission and also associated with regions in the final stages of the class~II maser phase.  Both class~I methanol and water masers are collisionally pumped and can be excited by outflows, expansion shocks and similar phenomena.  \citet{Breen+14a,Titmarsh+14,Titmarsh+16} suggested that this made water masers a less discriminatory probe of evolutionary status, as the presence (or absence) of the water maser emission is governed by the interaction of outflows (and/or shocks) with surrounding molecular gas in the star formation region.  This explanation appears a reasonable one for the class~I methanol masers as well, with class~I methanol masers arising where ever low-velocity shocks interact with molecular gas with sufficient methanol abundance (noting that such shocks themselves tend to enhance methanol abundance through heating and evaporation of ices from dust grain mantles).  In this scenario the class~I methanol maser phase traces the duration over which a star formation region has sufficiently high methanol abundance - from the release of methanol into the gas-phase from icy grain mantles, to the eventual destruction of the gas-phase methanol through a combination of chemical processing and dissociation from high-energy radiation or shocks.  Whereas the class~II methanol maser phase requires both the presence of gas-phase methanol and appropriate physical conditions (temperature, density etc) in the gas close to the young star.  Hence, we expect the radiatively-pumped OH and class~II methanol masers to be much more directly influenced by the young high-mass star and hence a better tracer of evolutionary phase  than the collisionally pumped water and class~I methanol masers.

\section{Conclusions}

We have made the first high resolution observations of the 37.7-, 38.3- and 38.5-GHz class~II methanol maser transitions towards a sample of eleven sources.  We find that to within the absolute astrometric accuracy (around 0.4 arcseconds), these class~II transitions arise from the same location as the strong 6.7-GHz class~II methanol masers.  The spectral and imaging data both suggest that the 37.7-GHz masers are less co-spatial with the 6.7-GHz maser spots than has previously been observed for the 12.2-GHz class~II transition.  The peak velocity of the 37.7-GHz masers is found to be preferentially blueshifted with respect to the systemic velocity, suggesting that these maser arise in molecular gas which is outflowing from the region in the general direction of the observer.  The detection of class~I methanol masers towards what are hypothesised to be star formation regions towards the end of the class~II methanol maser phase, combined with the results of recent untargetted searches for class~I methanol masers suggest that the duration of the  class~I methanol maser phase exceeds that of class~II methanol masers.

\section*{Acknowledgements}

The Australia Telescope Compact Array is part of the Australia Telescope which is funded by the Commonwealth of Australia for operation as a National Facility managed by CSIRO.  AMS was supported by the Russian Science Foundation (grant 18-12-00193).  This research has made use of NASA's Astrophysics Data System Abstract Service.  This work makes use of observations made with the Spitzer Space Telescope, which is operated by the Jet Propulsion Laboratory, California Institute of Technology under a contract with NASA. This publication makes use of data products from the Wide-field Infrared Survey Explorer, which is a joint project of the University of California, Los Angeles, and the Jet Propulsion Laboratory/California Institute of Technology, funded by the National Aeronautics and Space Administration.  The ATLASGAL project is a collaboration between the Max-Planck-Gesellschaft, the European Southern Observatory (ESO) and the Universidad de Chile. It includes projects E-181.C-0885, E-078.F-9040(A), M-079.C-9501(A), M-081.C-9501(A) plus Chilean data.

\bibliography{/Users/sellings/tex/references.bib}

\appendix

\section{Environment Figures} \label{sec:environ}

\begin{figure*}
  \psfig{file=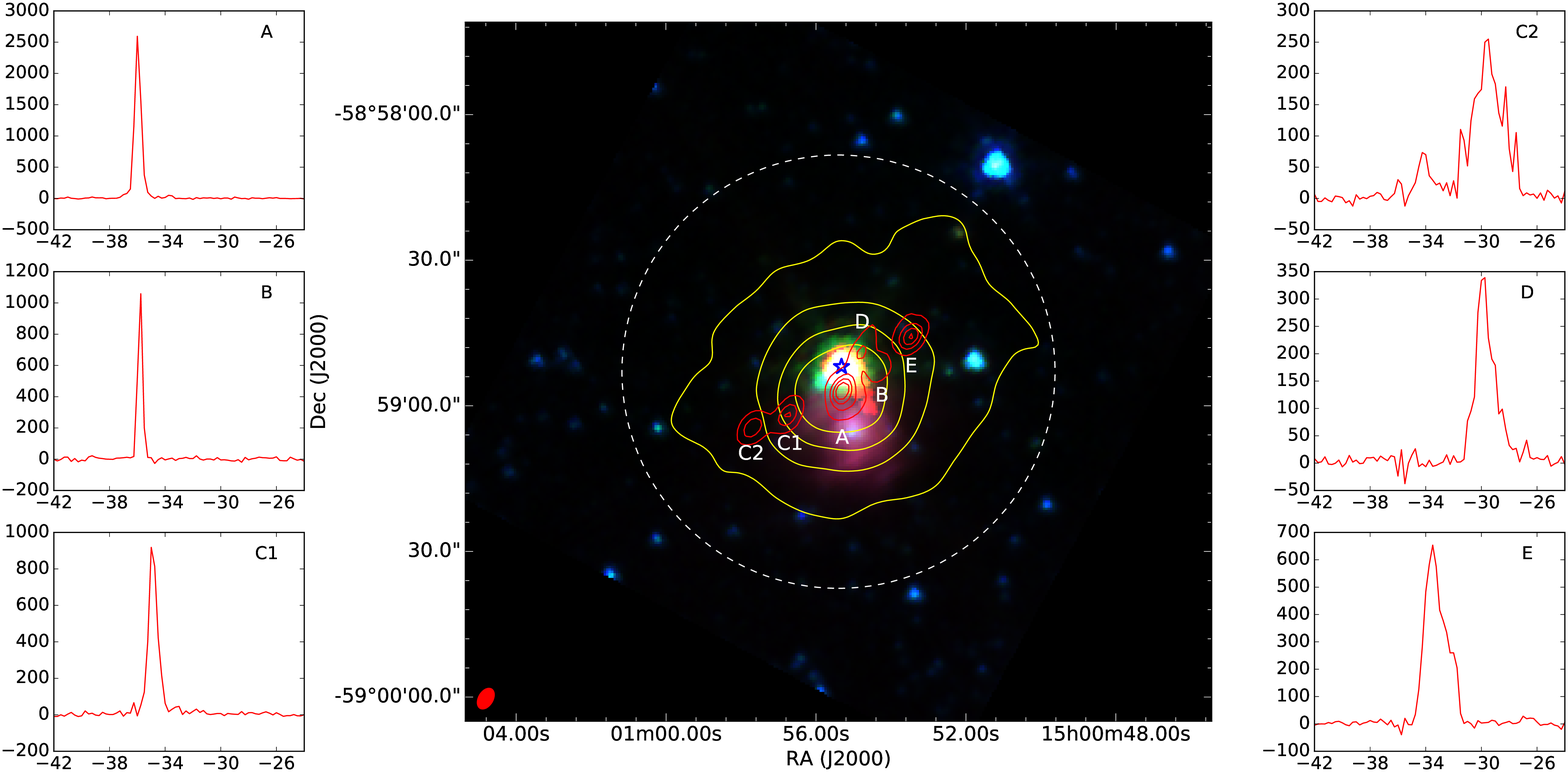,width=0.90\textwidth}
  \caption{The environment of the class~II methanol maser G\,318.948$-$0.196.  The background image is from {\em Spitzer} IRAC observations from the GLIMPSE survey \citep{Benjamin+03} with blue, green, and red from the 3.6, 4.5, and 8.0 $\mu$m bands, respectively.  The blue star marks the location of the 37.7-GHz methanol masers determined from the current observations.  The red contours are the integrated 36.2-GHz methanol maser emission from the current ATCA observations with contours at 10, 30, 50 and 70 percent of 1.58 Jy \kms\/ beam$^{-1}$.  The yellow contours are the 870~$\mu$m from the ATLASGAL survey \citep{Schuller+09}, with contours at 10, 30, 50 and 70 percent of 7.41 Jy beam$^{-1}$.  The white-dashed circle shows the half-power point of the ATCA primary beam and the red-filled ellipse in the bottom-left corner represents the synthesised beam of the 36.2-GHz observations.  The spectra to the left and right of the image show the 36.2-GHz methanol maser emission extracted from the image cube at the specified locations (see Table~\ref{tab:components}).  The x-axis of the spectra are line-of-sight velocity in \kms\/ and the y-axis of the spectra show the flux density in mJy.}
  \label{fig:g318}
\end{figure*} 

\begin{figure*}
  \psfig{file=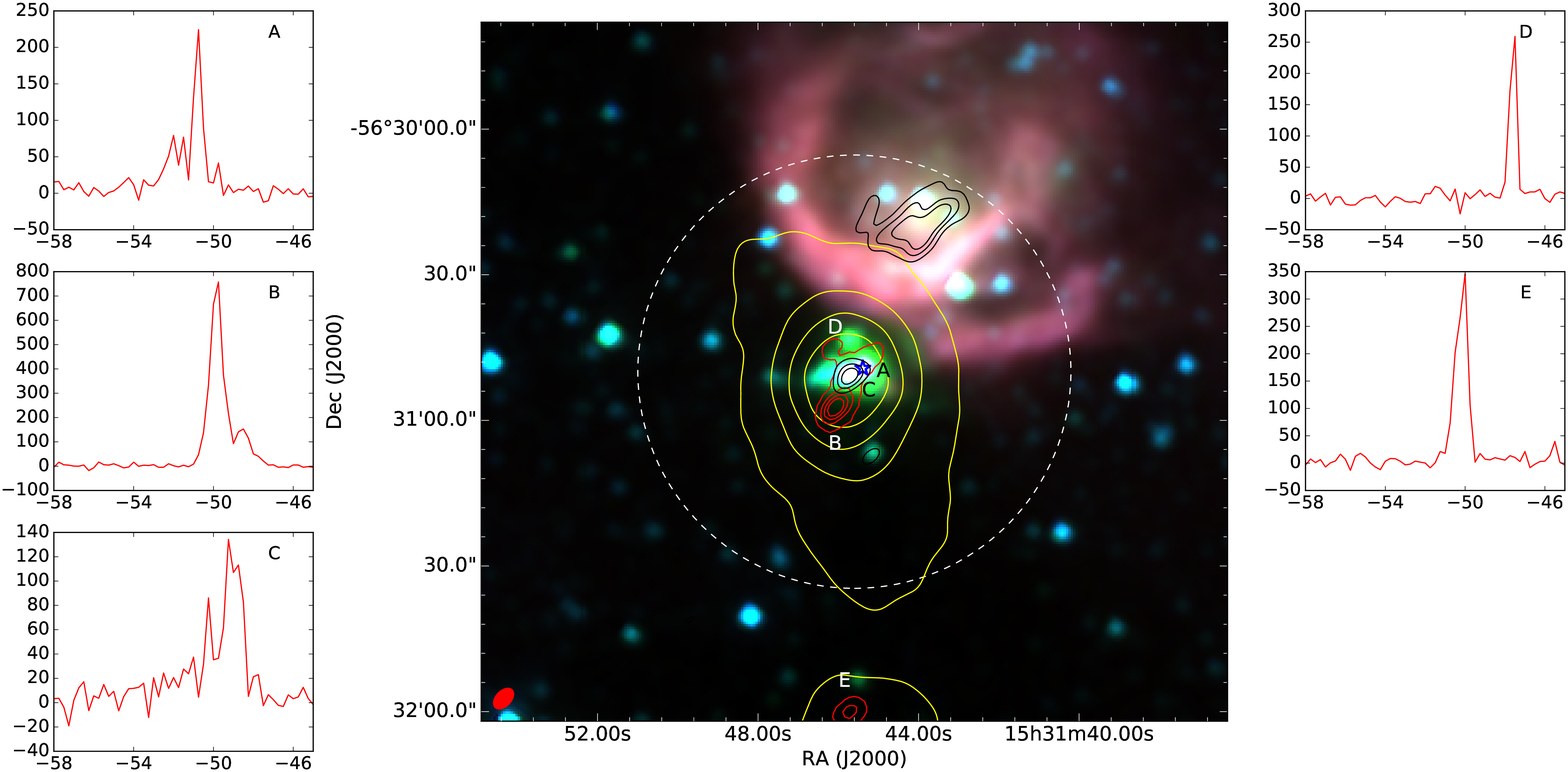,width=0.90\textwidth}
    \caption{The environment of the class~II methanol maser G\,323.740$-$0.263.  The background image is from {\em Spitzer} IRAC observations from the GLIMPSE survey \citep{Benjamin+03} with blue, green, and red from the 3.6, 4.5, and 8.0 $\mu$m bands, respectively.  The blue star marks the location of the 37.7-GHz methanol masers determined from the current observations.  The red contours are the integrated 36.2-GHz methanol maser emission from the current ATCA observations with contours at 10, 30, 50 and 70 percent of 0.81 Jy \kms\/ beam$^{-1}$.  The black contours are the 8-mm radio continuum emission from the current ATCA observations, with contours at 30, 50 and 70 percent of 0.0028 Jy beam$^{-1}$.The yellow contours are the 870~$\mu$m from the ATLASGAL survey \citep{Schuller+09}, with contours at 10, 30, 50 and 70 percent of 10.47 Jy beam$^{-1}$.  The white-dashed circle shows the half-power point of the ATCA primary beam and the red-filled ellipse in the bottom-left corner represents the synthesised beam of the 36.2-GHz observations.  The spectra to the left and right of the image show the 36.2-GHz methanol maser emission extracted from the image cube at the specified locations (see Table~\ref{tab:components}).  The x-axis of the spectra are line-of-sight velocity in \kms\/ and the y-axis of the spectra show the flux density in mJy.}
  \label{fig:g323}
\end{figure*} 

\begin{figure*}
  \psfig{file=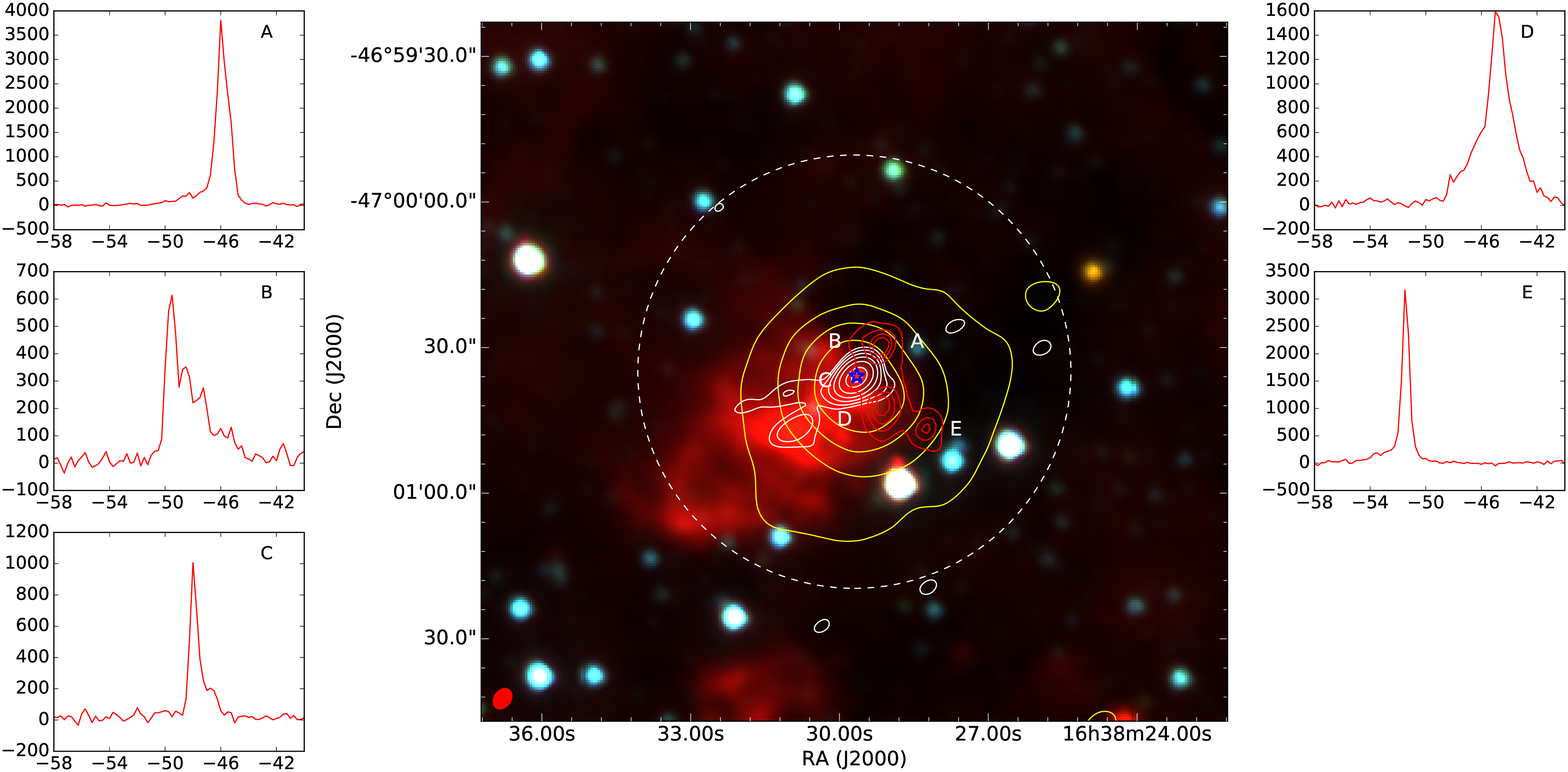,width=0.90\textwidth}
    \caption{The environment of the class~II methanol maser G\,337.705$-$0.053.  The background image is from {\em Spitzer} IRAC observations from the GLIMPSE survey \citep{Benjamin+03} with blue, green, and red from the 3.6, 4.5, and 8.0 $\mu$m bands, respectively.  The blue star marks the location of the 37.7-GHz methanol masers determined from the current observations.  The red contours are the integrated 36.2-GHz methanol maser emission from the current ATCA observations with contours at 10, 30, 50 and 70 percent of 5.05 Jy \kms\/ beam$^{-1}$.  The white contours are the 8-mm radio continuum emission from the current ATCA observations, with contours at 2, 4, 8, 16, 32, 64 and 90 percent of 0.32 Jy beam$^{-1}$.  The yellow contours are the 870~$\mu$m from the ATLASGAL survey \citep{Schuller+09}, with contours at 10, 30, 50 and 70 percent of 13.97 Jy beam$^{-1}$.  The white-dashed circle shows the half-power point of the ATCA primary beam and the red-filled ellipse in the bottom-left corner represents the synthesised beam of the 36.2-GHz observations.  The spectra to the left and right of the image show the 36.2-GHz methanol maser emission extracted from the image cube at the specified locations (see Table~\ref{tab:components}).  The x-axis of the spectra are line-of-sight velocity in \kms\/ and the y-axis of the spectra show the flux density in mJy.}
   \label{fig:g337}
\end{figure*} 

\begin{figure*}
  \psfig{file=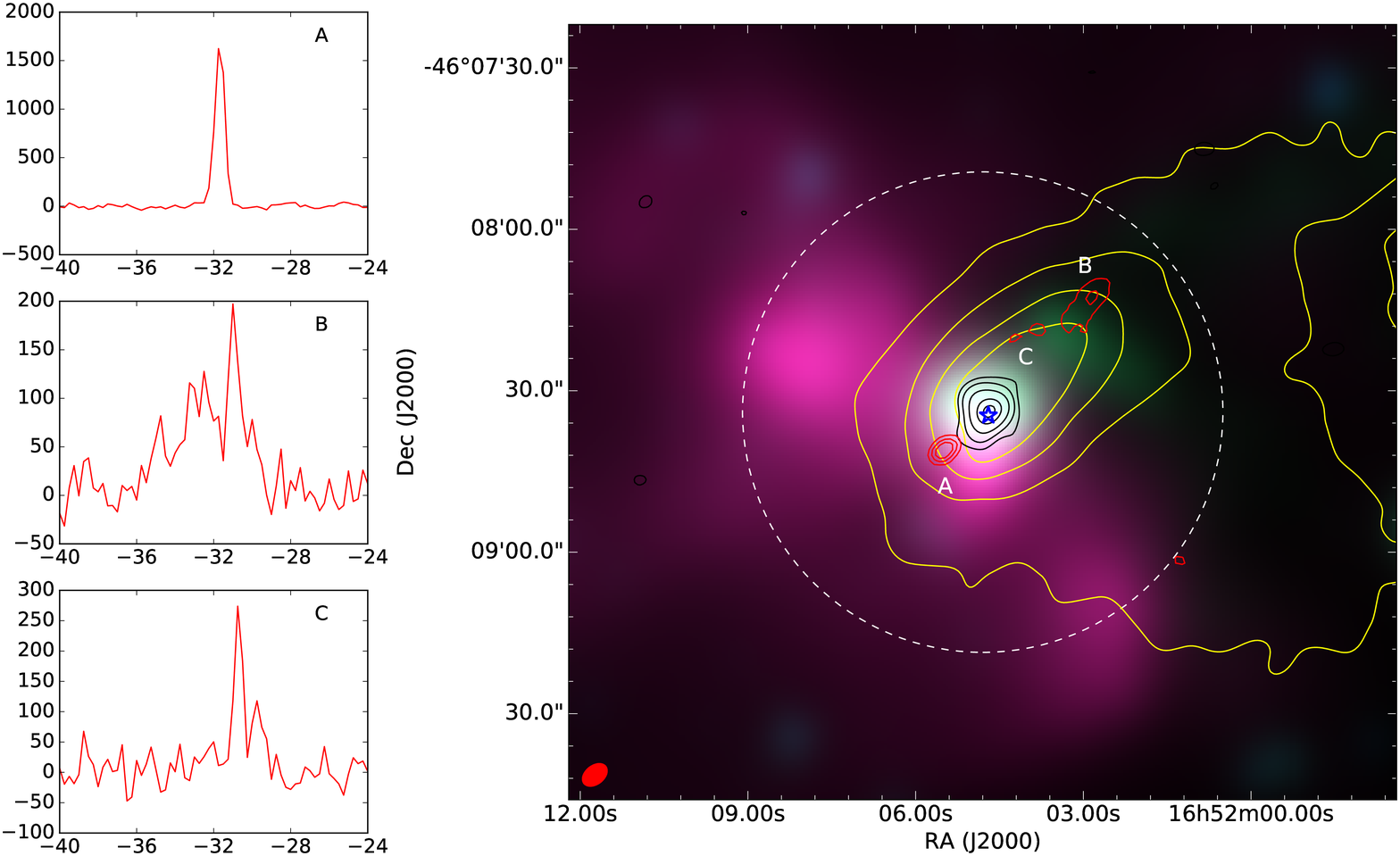,width=0.80\textwidth}
    \caption{The environment of the class~II methanol maser G\,339.884$-$1.259.  The background image is from {\em WISE} observations \citep{Wright+10} with blue, green, and red from the 3.4, 4.6, and 12 $\mu$m bands, respectively.  The blue star marks the location of the 37.7-GHz methanol masers determined from the current observations.  The red contours are the integrated 36.2-GHz methanol maser emission from the current ATCA observations with contours at 30, 50 and 70 percent of 1.06 Jy \kms\/ beam$^{-1}$.  The black contours are the 8-mm radio continuum emission from the current ATCA observations, with contours at 8, 16, 32, 64 and 90 percent of 0.008 Jy beam$^{-1}$.  The yellow contours are the 870~$\mu$m from the ATLASGAL survey \citep{Schuller+09}, with contours at 10, 30, 50 and 70 percent of 8.60 Jy beam$^{-1}$.  The white-dashed circle shows the half-power point of the ATCA primary beam and the red-filled ellipse in the bottom-left corner represents the synthesised beam of the 36.2-GHz observations.  The spectra to the left of the image show the 36.2-GHz methanol maser emission extracted from the image cube at the specified locations (see Table~\ref{tab:components}).  The x-axis of the spectra are line-of-sight velocity in \kms\/ and the y-axis of the spectra show the flux density in mJy.}
  \label{fig:g339}
\end{figure*} 

\begin{figure*}
  \psfig{file=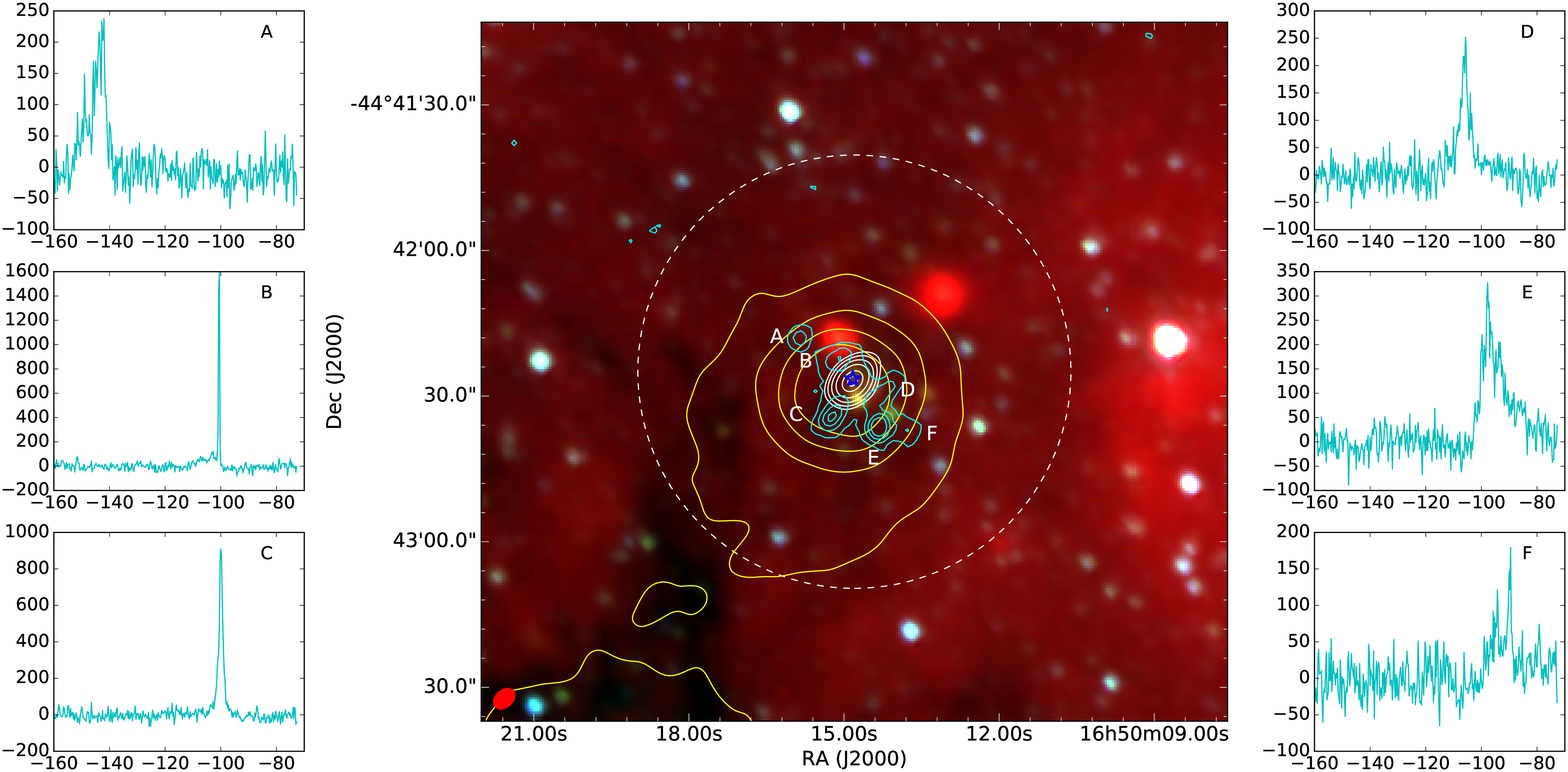,width=0.90\textwidth}
    \caption{The environment of the class~II methanol maser G\,340.785$-$0.096.  The background image is from {\em Spitzer} IRAC observations from the GLIMPSE survey \citep{Benjamin+03} with blue, green, and red from the 3.6, 4.5, and 8.0 $\mu$m bands, respectively.  The blue star marks the location of the 37.7-GHz methanol masers determined from the current observations.  The cyan contours are the integrated 36.2-GHz methanol maser emission from the current ATCA observations with contours at 10, 30, 50 and 70 percent of 2.25 Jy \kms\/ beam$^{-1}$.  The white contours are the 8-mm radio continuum emission from the current ATCA observations, with contours at 4, 8, 16, 32, 64 and 90 percent of 0.043 Jy beam$^{-1}$.  The yellow contours are the 870~$\mu$m from the ATLASGAL survey \citep{Schuller+09}, with contours at 10, 30, 50 and 70 percent of 4.09 Jy beam$^{-1}$.  The white-dashed circle shows the half-power point of the ATCA primary beam and the red-filled ellipse in the bottom-left corner represents the synthesised beam of the 36.2-GHz observations.  The spectra to the left and right of the image show the 36.2-GHz methanol maser emission extracted from the image cube at the specified locations (see Table~\ref{tab:components}).  The x-axis of the spectra are line-of-sight velocity in \kms\/ and the y-axis of the spectra show the flux density in mJy.}  \label{fig:g340}
\end{figure*} 

%\begin{figure*}
%  \psfig{file=eps/g345_all.eps,width=0.90\textwidth}
%    \caption{The environment of the class~II methanol maser G\,345.010$+$1.792.  The background image is from {\em WISE} observations \citep{Wright+10} with blue, green, and red from the 3.4, 4.6, and 12 $\mu$m bands, respectively.  The blue star marks the location of the 37.7-GHz methanol masers determined from the current observations.  The red contours are the integrated 36.2-GHz methanol maser emission from the current ATCA observations with contours at 4, 8, 16, 32, 64 and 90 percent of 6.77 Jy \kms\/ beam$^{-1}$.  The black contours are the 8-mm radio continuum emission from the current ATCA observations, with contours at 2, 4, 8, 16, 32, 64 and 90 percent of 0.43 Jy beam$^{-1}$.  The white-dashed circle shows the half-power point of the ATCA primary beam and the red-filled ellipse in the bottom-left corner represents the synthesised beam of the 36.2-GHz observations.  The spectra to the left and right of the image show the 36.2-GHz methanol maser emission extracted from the image cube at the specified locations (see Table~\ref{tab:components}).  The x-axis of the spectra are line-of-sight velocity in \kms\/ and the y-axis of the spectra show the flux density in mJy.}
%  \label{fig:g345}
%\end{figure*} 

\begin{figure*}
  \psfig{file=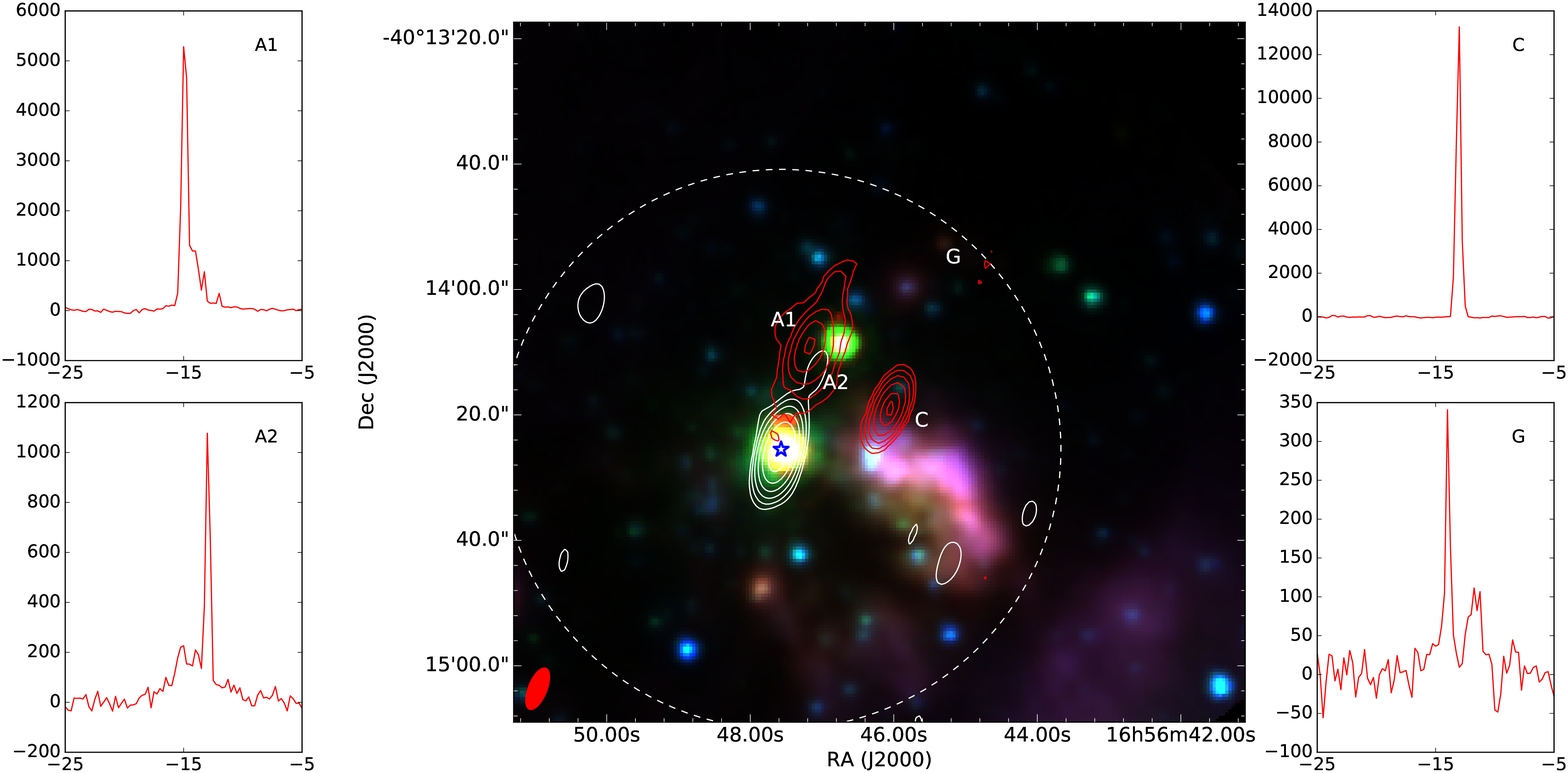,width=0.90\textwidth}
    \caption{The environment of the class~II methanol maser G\,345.010$+$1.792.  The background image is from {\em Spitzer} IRAC observations with blue, green, and red from the 3.6, 4.5, and 8.0 $\mu$m bands, respectively. .  The blue star marks the location of the 37.7-GHz methanol masers determined from the current observations.  The red contours are the integrated 36.2-GHz methanol maser emission from the current ATCA observations with contours at 4, 8, 16, 32, 64 and 90 percent of 6.77 Jy \kms\/ beam$^{-1}$.  The white contours are the 8-mm radio continuum emission from the current ATCA observations, with contours at 2, 4, 8, 16, 32, 64 and 90 percent of 0.43 Jy beam$^{-1}$.  The white-dashed circle shows the half-power point of the ATCA primary beam and the red-filled ellipse in the bottom-left corner represents the synthesised beam of the 36.2-GHz observations.  The spectra to the left and right of the image show the 36.2-GHz methanol maser emission extracted from the image cube at the specified locations (see Table~\ref{tab:components}).  The x-axis of the spectra are line-of-sight velocity in \kms\/ and the y-axis of the spectra show the flux density in mJy.}
  \label{fig:g345}
\end{figure*} 

\begin{figure*}
  \psfig{file=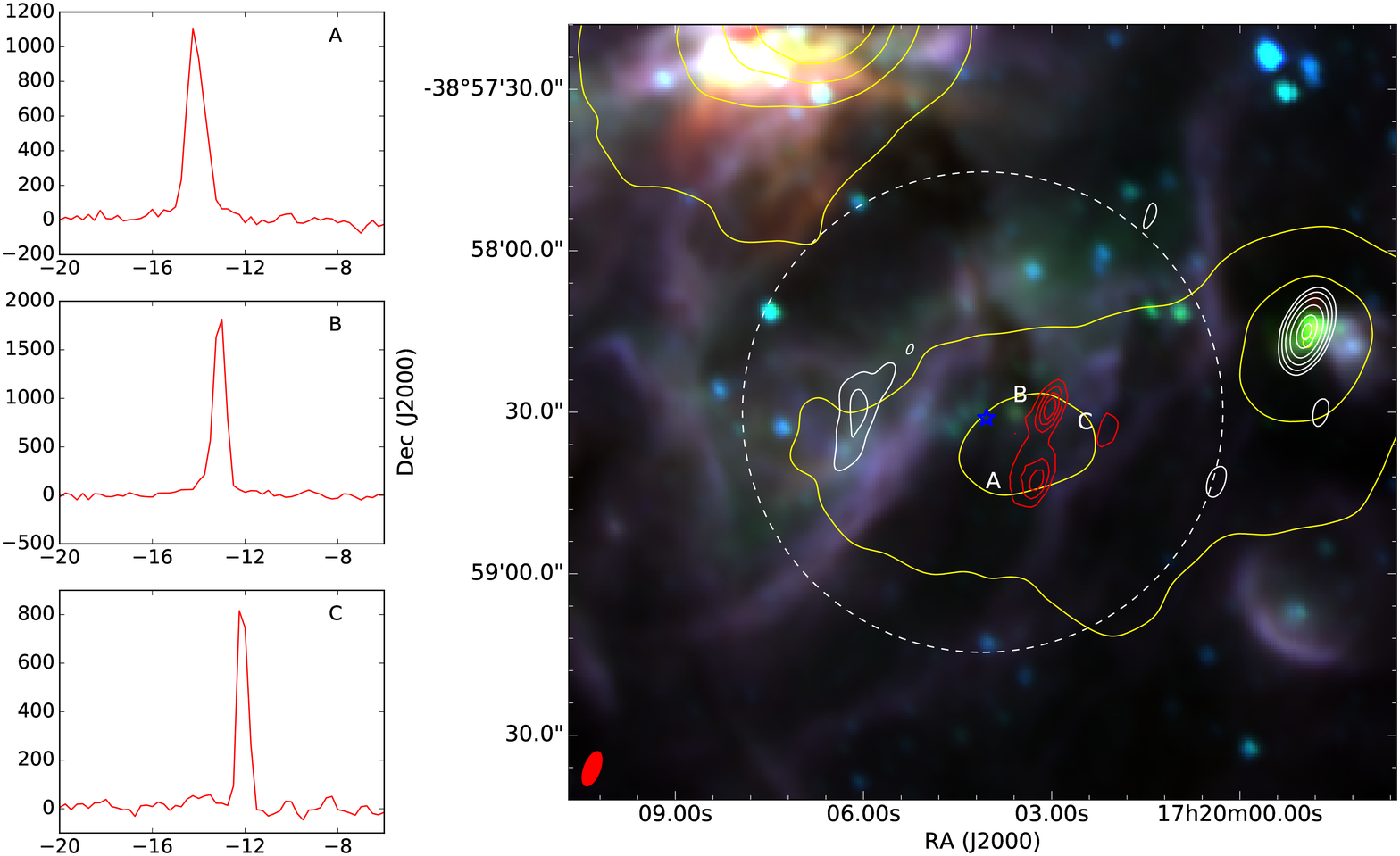,width=0.80\textwidth}
    \caption{The environment of the class~II methanol maser G\,348.703$-$1.043.  The background image is from {\em Spitzer} IRAC observations from the GLIMPSE survey \citep{Benjamin+03} with blue, green, and red from the 3.6, 4.5, and 8.0 $\mu$m bands, respectively.  The blue star marks the location of the 37.7-GHz methanol masers determined from the current observations.  The red contours are the integrated 36.2-GHz methanol maser emission from the current ATCA observations with contours at 20, 40, 60 and 80 percent of 1.44 Jy \kms\/ beam$^{-1}$.  The white contours are the 8-mm radio continuum emission from the current ATCA observations, with contours at 4, 8, 16, 32, 64 and 90 percent of 0.34 Jy beam$^{-1}$.  The yellow contours are the 870~$\mu$m from the ATLASGAL survey \citep{Schuller+09}, with contours at 10, 30, 50 and 70 percent of 24.1 Jy beam$^{-1}$.  The white-dashed circle shows the half-power point of the ATCA primary beam and the red-filled ellipse in the bottom-left corner represents the synthesised beam of the 36.2-GHz observations.  The spectra to the left of the image show the 36.2-GHz methanol maser emission extracted from the image cube at the specified locations (see Table~\ref{tab:components}).  The x-axis of the spectra are line-of-sight velocity in \kms\/ and the y-axis of the spectra show the flux density in mJy.}    \label{fig:g348}
\end{figure*} 

\begin{figure*}
  \psfig{file=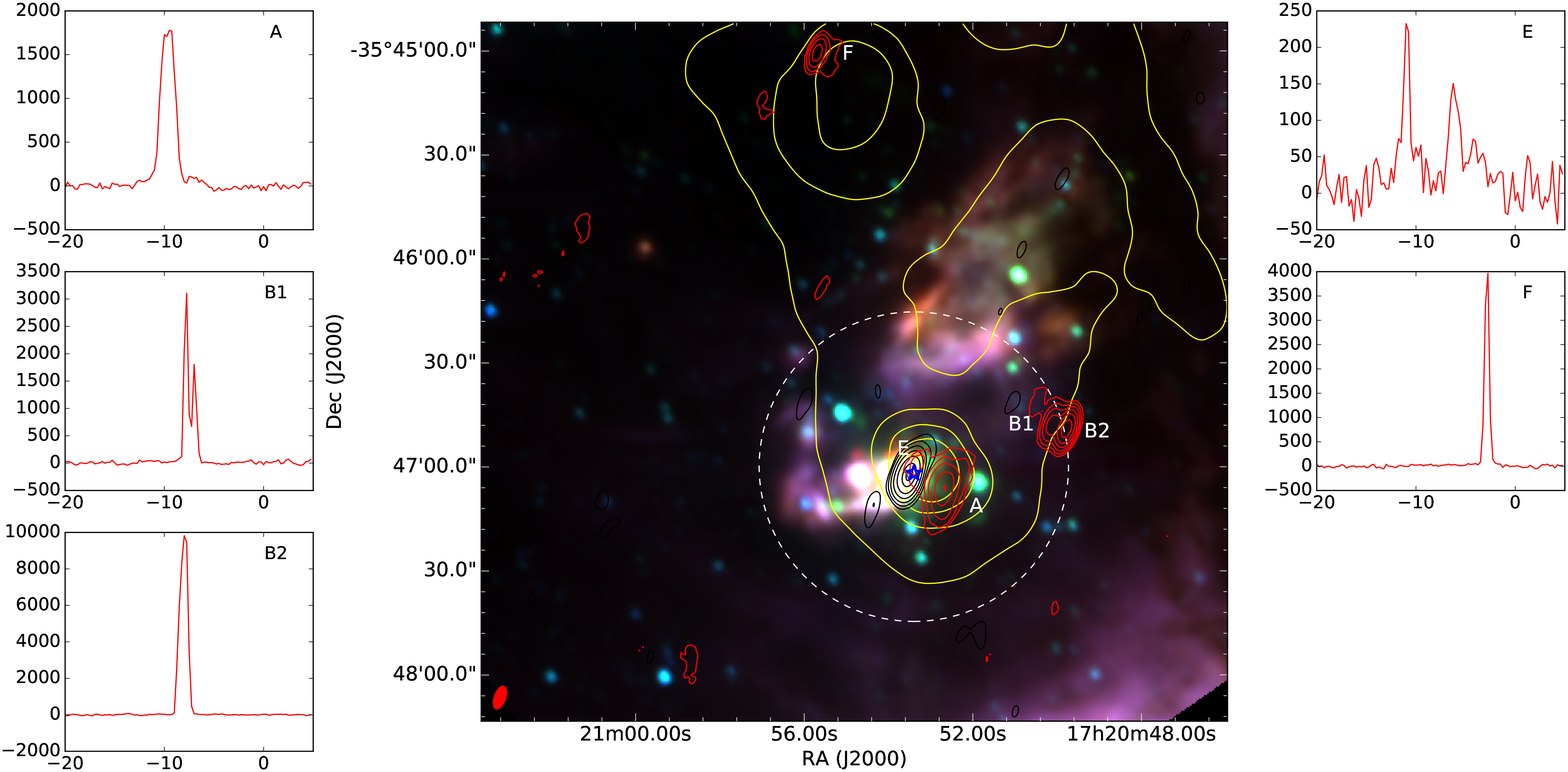,width=0.90\textwidth}
    \caption{The environment of the class~II methanol maser NGC6334F.  The background image is from {\em Spitzer} IRAC observations from the GLIMPSE survey \citep{Benjamin+03} with blue, green, and red from the 3.6, 4.5, and 8.0 $\mu$m bands, respectively.  The blue star marks the location of the 37.7-GHz methanol masers determined from the current observations.  The red contours are the integrated 36.2-GHz methanol maser emission from the current ATCA observations with contours at 2, 4, 8, 16, 32 and 64 percent of 10.0 Jy \kms\/ beam$^{-1}$.  The black contours are the 8-mm radio continuum emission from the current ATCA observations, with contours at 2, 4, 8, 16, 32, 64 and 90 percent of 2.01 Jy beam$^{-1}$.  The yellow contours are the 870~$\mu$m from the ATLASGAL survey \citep{Schuller+09}, with contours at 10, 30, 50 and 70 percent of 59.3 Jy beam$^{-1}$.  The white-dashed circle shows the half-power point of the ATCA primary beam and the red-filled ellipse in the bottom-left corner represents the synthesised beam of the 36.2-GHz observations.  The spectra to the left and right of the image show the 36.2-GHz methanol maser emission extracted from the image cube at the specified locations (see Table~\ref{tab:components}).  The x-axis of the spectra are line-of-sight velocity in \kms\/ and the y-axis of the spectra show the flux density in mJy.}      \label{fig:ngc6334f}
\end{figure*} 

\begin{figure*}
  \psfig{file=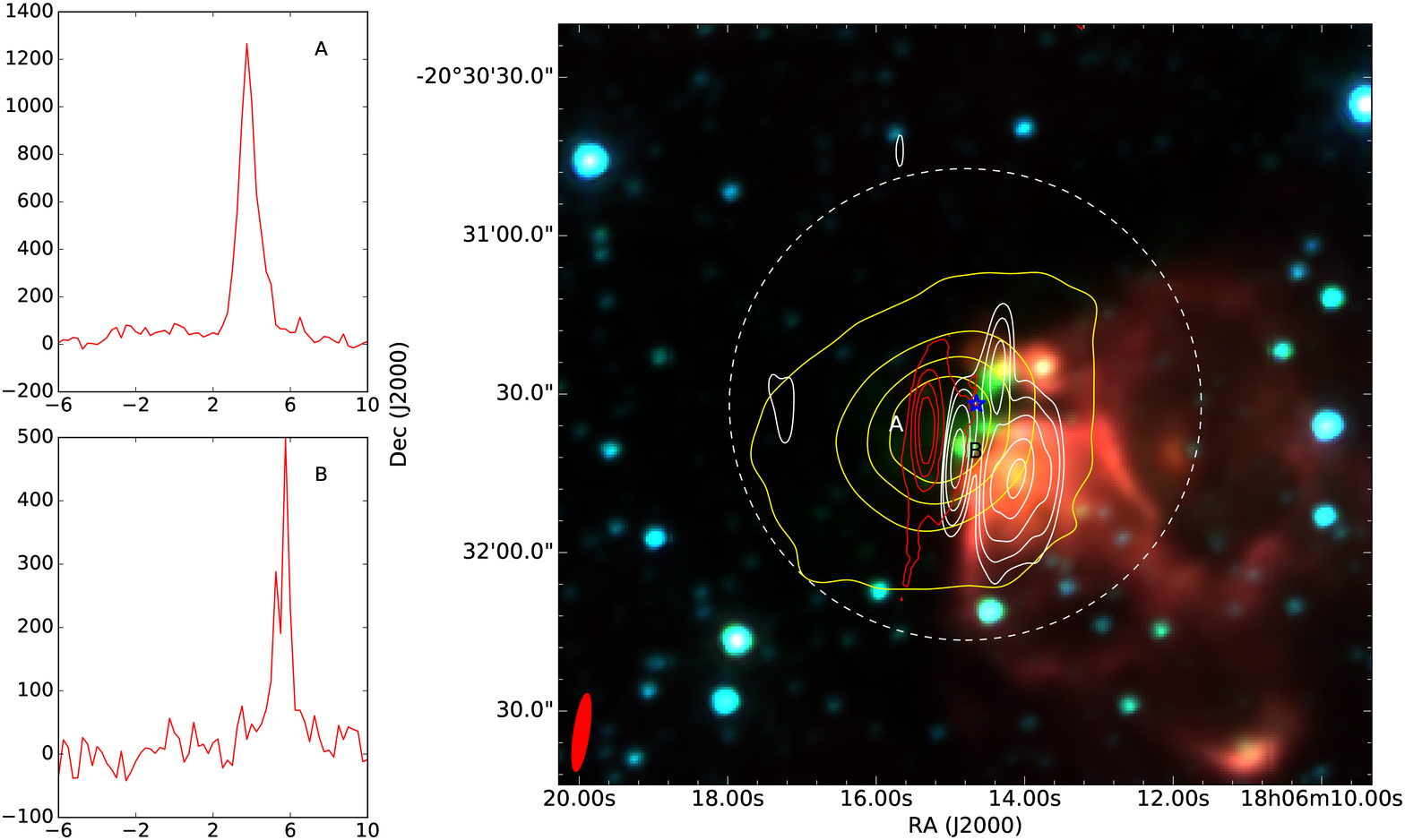,width=0.80\textwidth}
    \caption{The environment of the class~II methanol maser G\,9.621$+$0.196.  The background image is from {\em Spitzer} IRAC observations from the GLIMPSE survey \citep{Benjamin+03} with blue, green, and red from the 3.6, 4.5, and 8.0 $\mu$m bands, respectively.  The blue star marks the location of the 37.7-GHz methanol masers determined from the current observations.  The red contours are the integrated 36.2-GHz methanol maser emission from the current ATCA observations with contours at 10, 30, 50 and 70 percent of 1.9 Jy \kms\/ beam$^{-1}$.  The white contours are the 8-mm radio continuum emission from the current ATCA observations, with contours at 8, 16, 32, 64 and 90 percent of 0.06 Jy beam$^{-1}$.  The yellow contours are the 870~$\mu$m from the ATLASGAL survey \citep{Schuller+09}, with contours at 10, 30, 50 and 70 percent of 12.5 Jy beam$^{-1}$.  The white-dashed circle shows the half-power point of the ATCA primary beam and the red-filled ellipse in the bottom-left corner represents the synthesised beam of the 36.2-GHz observations.  The spectra to the left of the image show the 36.2-GHz methanol maser emission extracted from the image cube at the specified locations (see Table~\ref{tab:components}).  The x-axis of the spectra are line-of-sight velocity in \kms\/ and the y-axis of the spectra show the flux density in mJy.}        \label{fig:g9}
\end{figure*} 

\begin{figure*}
  \psfig{file=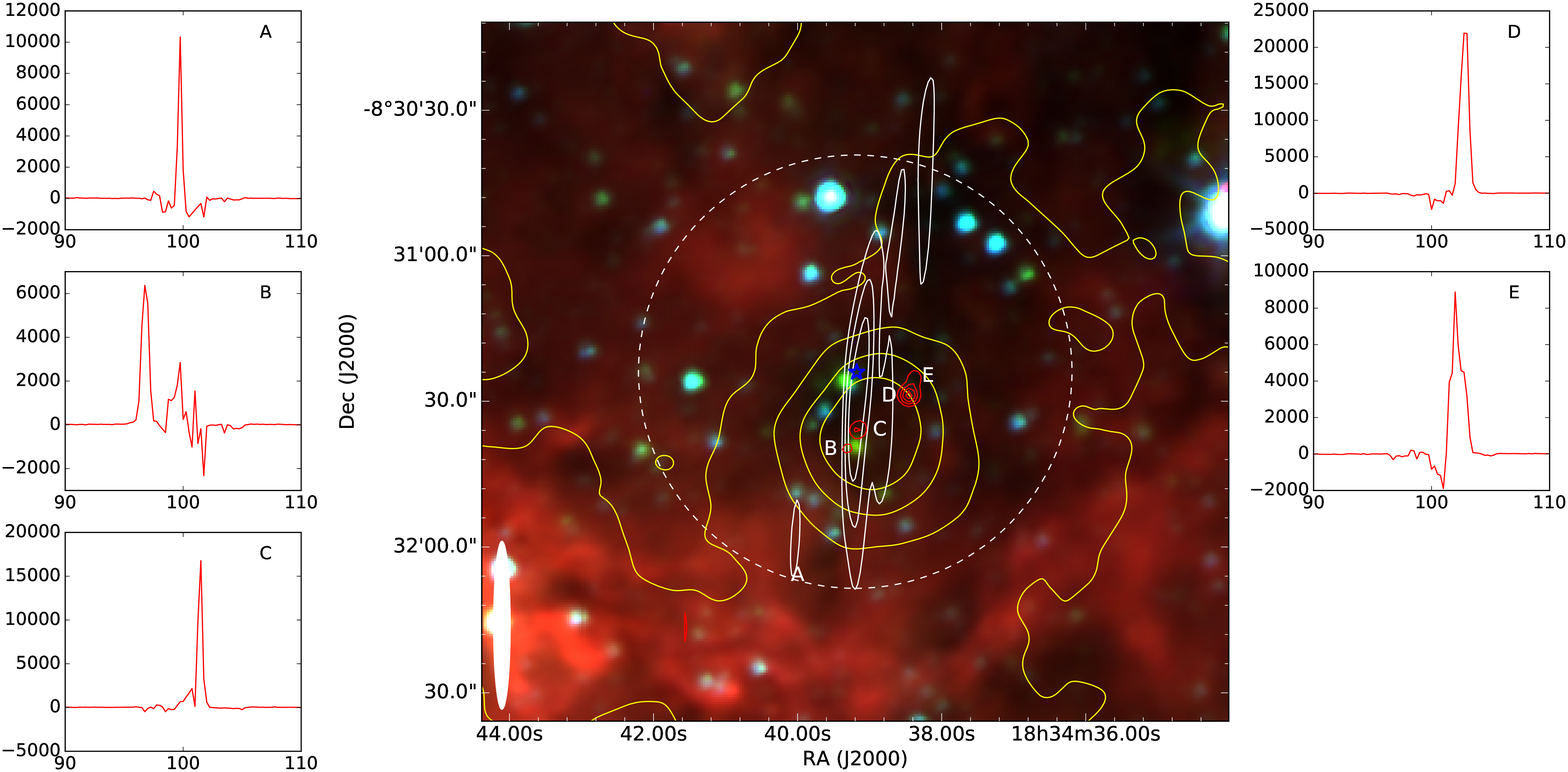,width=0.90\textwidth}
    \caption{The environment of the class~II methanol maser G\,23.440$-$0.182.  The background image is from {\em Spitzer} IRAC observations from the GLIMPSE survey \citep{Benjamin+03} with blue, green, and red from the 3.6, 4.5, and 8.0 $\mu$m bands, respectively.  The blue star marks the location of the 37.7-GHz methanol masers determined from the current observations.  The red contours are the integrated 36.2-GHz methanol maser emission from the current ATCA observations with contours at 25, 45, 65 and 85 percent of 18.1 Jy \kms\/ beam$^{-1}$.  The white contours are the 8-mm radio continuum emission from the current ATCA observations, with contours at 40, 60 and 80 percent of 0.0025 Jy beam$^{-1}$.  The yellow contours are the 870~$\mu$m from the ATLASGAL survey \citep{Schuller+09}, with contours at 10, 30, 50 and 70 percent of 6.85 Jy beam$^{-1}$.  The white-dashed circle shows the half-power point of the ATCA primary beam and the white-filled ellipse in the bottom-left corner represents the synthesised beam of the 8-mm continuum observations, a 3.2 arcsecond circular restoring beam was used for the 36.2-GHz observations.  The spectra to the left and right of the image show the 36.2-GHz methanol maser emission extracted from the image cube at the specified locations (see Table~\ref{tab:components}).  The x-axis of the spectra are line-of-sight velocity in \kms\/ and the y-axis of the spectra show the flux density in mJy.}       \label{fig:g23}
\end{figure*}

\begin{figure*}
  \psfig{file=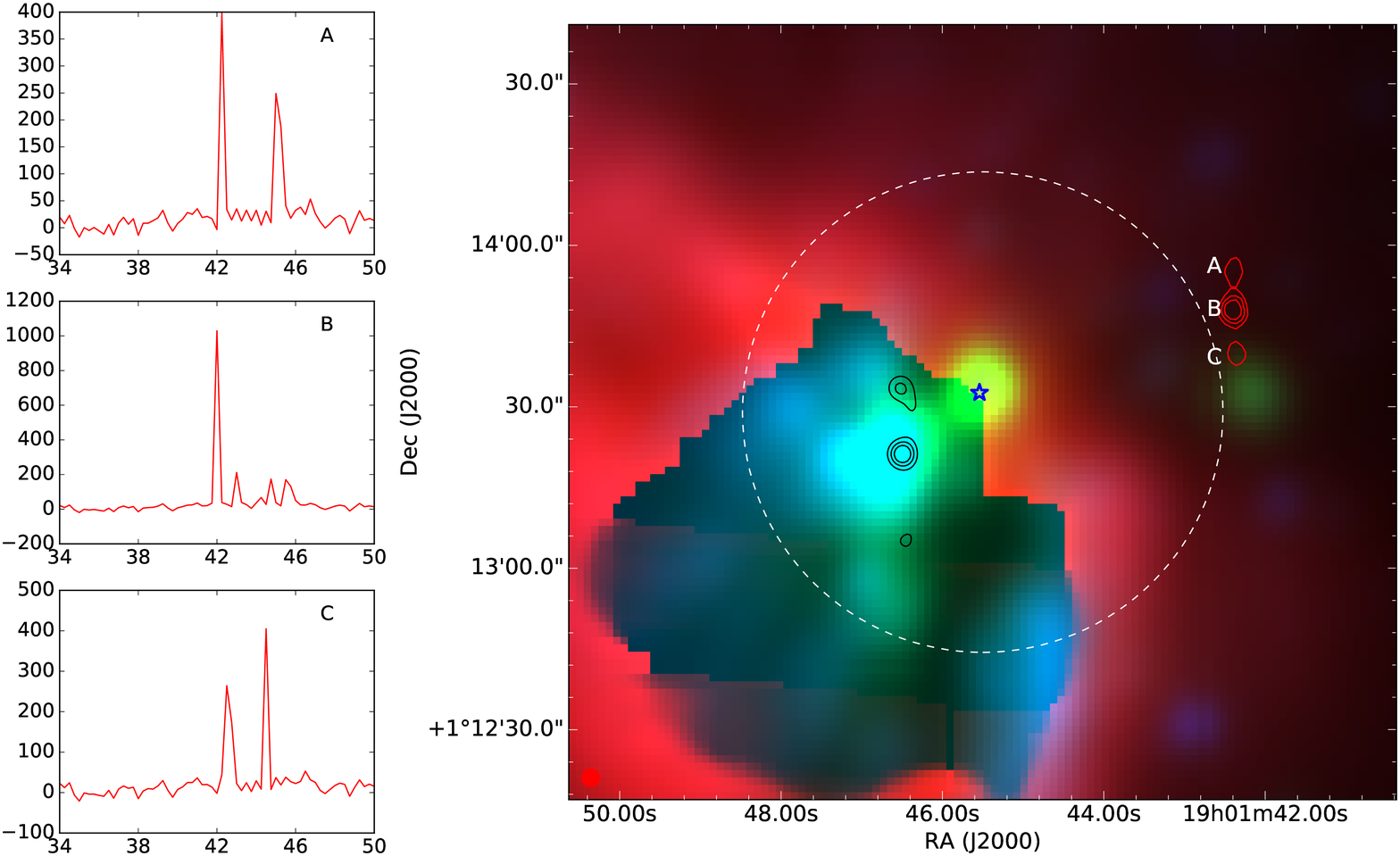,width=0.80\textwidth}
    \caption{The environment of the class~II methanol maser G\,35.201$-$1.736.  The background image is from {\em WISE} observations \citep{Wright+10} with blue, green, and red from the 3.4, 4.6, and 12 $\mu$m bands, respectively.  The blue star marks the location of the 6.7-GHz methanol masers \citep{Breen+15}.  The red contours are the integrated 36.2-GHz methanol maser emission from the current ATCA observations with contours at 30, 50 and 70 percent of 0.47 Jy \kms\/ beam$^{-1}$.  The black contours are the 8-mm radio continuum emission from the current ATCA observations, with contours at 30, 50 and 70 percent of 0.44 Jy beam$^{-1}$.  The white-dashed circle shows the half-power point of the ATCA primary beam and the red-filled ellipse in the bottom-left corner represents the circular 4 arcsecond restoring beam used for the 36.2-GHz and 8-mm continuum observations.  The spectra to the left of the image show the 36.2-GHz methanol maser emission extracted from the image cube at the specified locations (see Table~\ref{tab:components}).  The x-axis of the spectra are line-of-sight velocity in \kms\/ and the y-axis of the spectra show the flux density in mJy.}
  \label{fig:g35}
\end{figure*} 

\begin{table*}
\caption{36.2-GHz class~I methanol maser components.  The spectra shown at the sides of Figures~\ref{fig:g318} -- \ref{fig:g35} are extracted from the image cube at the location given (over the region of the synthesised beam).  Components in italics were excluded from the integrated spectra shown in Figure~\ref{fig:meth36} (usually because they are significantly offset from the bulk of the emission).  Components marked with a $^*$ are additional components to those in \citet{Voronkov+14} (for the five sources which overlap).}
 \begin{tabular}{lcllrrcl} \hline
      \multicolumn{1}{c}{\bf Source} & \multicolumn{1}{c}{\bf Component} & \multicolumn{1}{c}{\bf Right Ascension}  & \multicolumn{1}{c}{\bf Declination} & \multicolumn{1}{c}{\bf Peak Flux} & \multicolumn{3}{c}{\bf Velocity}  \\
            \multicolumn{1}{c}{\bf name}    &  & \multicolumn{1}{c}{\bf (J2000)} & \multicolumn{1}{c}{\bf (J2000)} &  \multicolumn{1}{c}{\bf Density} & \multicolumn{3}{c}{\bf Range} \\
          &  & \multicolumn{1}{c}{\bf $h$~~~$m$~~~$s$}& \multicolumn{1}{c}{\bf $^\circ$~~~$\prime$~~~$\prime\prime$} & \multicolumn{1}{c}{\bf (Jy)} & \multicolumn{3}{c}{\bf (\kms)} \\ \hline \hline
 G\,318.948$-$0.196 & A             & 15:00:55.27 & $-$58:58:57.0 & 2.6 & -36.2 & -- &  -35.5 \\
                                  & B             & 15:00:54.24 & $-$58:58:52.0 & 1.1 & -36.0 & -- & -35.5 \\
                                  & C1$^{*}$ & 15:00:56.69 & $-$58:59:02.0 & 0.9 & -35.2 & -- & -34.2 \\
                                  & C2$^{*}$ & 15:00:57.73 & $-$58:59:04.0 & 0.26 & -34.5 & -- &  -27.5 \\
                                  & D             & 15:00:54.77 & $-$58:58:49.0 & 0.34 & -31.0 & -- & -28.2 \\
                                  & E             & 15:00:53.46 & $-$58:58:46.0 & 0.65 & -34.5 & -- & -31.4 \\
 G\,323.740$-$0.263 & A             & 15:31:45.24 & $-$56:30:48.0 & 0.22 & -52.5& -- &-50.5 \\
                                  & B             & 15:31:46.08 & $-$56:30:58.0 & 0.76 & -50.5& -- &-48.0 \\
                                  & C             & 15:31:45.72 & $-$56:30:52.0 & 0.13 & -50.2& -- &-48.5 \\
                                  & D             & 15:31:46.08 & $-$56:30:44.0 & 0.26 & -47.7& -- &-47.5 \\
                                  & {\em E}$^*$ & 15:31:45.84 & $-$56:31:32.0 & 0.35 & -50.8& -- &-49.7 \\ % Not in Voronkov
 G\,337.705$-$0.053 & A    & 16:38:29.11 & $-$47:00:29.0           & 3.8 & -48.3& -- &-44.7 \\ % New source
                                  & B    & 16:38:29.51 & $-$47:00:28.0          & 0.61 & -50.0& -- &-45.2 \\
                                  & C    & 16:38:29.60 & $-$47:00:36.0          & 1.0 & -48.5& -- &-46.2 \\
                                  & D    & 16:38:29.11 & $-$47:00:42.0          & 1.6 & -48.3& -- &-41.7 \\
                                  & E    & 16:38:28.23 & $-$47:00:47.0    & 3.2 & -54.0& -- &-50.5 \\
 G\,339.884$-$1.259 & A     & 16:52:05.47 & $-$46:08:41.0 & 1.6 & -32.3& -- &-31.2 \\
                                  & B     & 16:52:02.88 & $-$46:08:13.0 & 0.20 & -33.2& -- &-30.7 \\
                                  & C$^{*}$ & 16:52:04.13 & $-$46:08:19.0 & 0.27 & -31.0& -- &-29.7 \\
G\,340.785$-$0.096 & A      & 16:50:15.83 & $-$44:42:18.0 & 0.24 & -151.5& -- &-141.0 \\
                                 & B      & 16:50:15.08 & $-$44:42:22.0 & 1.6 & -103.3& -- &-100.3 \\
                                 & C      & 16:50:15.27 & $-$44:42:34.0 & 0.91 & -101.8& -- &-98.5 \\
                                 & D      & 16:50:14.14 & $-$44:42:28.0 & 0.25 & -107.5& -- &-103.5 \\
                                 & E      & 16:50:14.33 & $-$44:42:36.0 & 0.33 & -100.3& -- &-88.3 \\
                                 & F      & 16:50:13.67 & $-$44:42:38.0 & 0.18 & -94.3& -- &-89.5 \\
G\,345.010$+$1.792 & A1$^*$ & 16:56:47.18 & $-$40:14:09.0 & 5.3 & -16.0& -- &-12.0 \\
                                  & A2$^*$ & 16:56:47.00 & $-$40:14:13.0 & 1.1 & -15.7& -- &-12.8 \\ 
                                  & C          &  16:56:46.04 & $-$40:14:19.0 & 13.3 & -13.5 & -- &-12.5 \\
                                  & G          &  16:56:44.64 & $-$40:13:55.0 & 0.34 & -14.3 & -- &-11.3 \\
G\,348.703$-$1.043 & A      &17:20:03.24 & $-$38:58:43.0 & 1.1 & -14.7& -- &-13.3 \\
                                 & B      &17:20:03.07 & $-$38:58:29.0 & 1.8 & -14.0& -- &-12.8 \\
                                 & C     & 17:20:02.13 & $-$38:58:34.0 & 0.82 & -12.2& -- &-11.8 \\
NGC6334F               & A     & 17:20:52.66 & $-$35:47:06.0  & 1.8 & -10.7& -- &-8.5 \\
                                 & B1$^*$ & 17:20:50.11 & $-$35:46:48.0 & 3.1 & -8.0& -- &-6.8 \\
                                 & B2$^*$ & 17:20:49.78 & $-$35:46:49.0 & 9.8 & -8.8& -- &-7.2 \\
                                 & E          & 17:20:53.24 & $-$35:47:00.0 & 0.23 & -11.3& -- &-5.8 \\ 
                                 & {\em F}$^*$ & 17:20:55.70 & $-$35:45:01.0 & 4.0 & -3.3& -- &-2.2 \\
G\,9.621$+$0.196    & A        & 18:06:15.30 & $-$20:31:39.0 & 1.3 & 3.0 & -- & 5.0\\
                                 & B        & 18:06:15.01 & $-$20:31:40.0 & 0.5 & 5.3& -- &6.0 \\
G\,23.440$-$0.182 & A          & 18:34:39.74 & $-$08:32:05.0 & 10.3 & 97.5& -- &100.0 \\
                               & B         & 18:34:39.34 & $-$08:31:40.0 & 5.6 & 95.7& -- &101.0 \\
                               & C         & 18:34:39.20 & $-$08:31:36.0 & 16.8 & 97.7& -- &102.0 \\
                               & D         & 18:34:38.46 & $-$08:31:29.0 & 22.0 & 101.2& -- &103.7 \\
                               & E         & 18:34:38.39 & $-$08:31:25.0 & 8.9 & 98.2& -- &103.2 \\
G\,35.201$-$1.736 & A         & 19:01:42.37 & $+$01:13:55.0 & 0.40 & 42.3& -- &45.3 \\
                               & B         & 19:01:42.37 & $+$01:13:48.0 & 1.0 & 42.0& -- &45.8 \\
                               & C         & 19:01:42.37 & $+$01:13:40.0 & 0.4 & 42.5& -- &44.5 \\ \hline 
\end{tabular} \label{tab:components}
\end{table*}

\end{document}